\documentclass[sort&compress,twocolumn,5p]{elsarticle} 

\usepackage{amsmath,amssymb}

\usepackage{color}

\usepackage[pdftex,breaklinks,
            pdftitle={Transmission Spectroscopy with the ACE-FTS Infrared Spectral Atlas of Earth: A Model Validation and Feasibility Study},
            pdfsubject={Extrasolar planets, atmospheric spectroscopy, remote sensing},
            pdfkeywords={Radiative transfer;  Planets and satellites: atmospheres;  Techniques: spectroscopic; ­Infrared: planetary systems},
            pdfauthor={S. Staedt and F. Schreier}]{hyperref}

\newcommand{\qeq}[1]  {Eq.~(\ref{#1})}
\newcommand{\qeqs}[1] {Eqs.~(\ref{#1})}
\newcommand{\qufig}[1] {Fig.~\ref{#1}}
\newcommand{\qutab}[1] {Tab.~\ref{#1}}
\newcommand{\D}{\mathrm{d}}
\newcommand{\E}{\mathrm{e}}

\newcommand{\T} {{\mathcal T}}
\newcommand{\cm}  {\ensuremath{\rm\, cm^{-1}}}
\newcommand{\mue} {\ensuremath{\rm\, \mu m}}

\newcommand{\chem}[1] {{\ensuremath{\mathrm{#1}}}}
\newcommand{\sinc} {\mathop{\rm sinc}}
\newcommand{\TO} {\ensuremath{\,\mbox{--}\,}}

\batchmode

\journal{Molecular Astrophysics, ~ received 25 July 2017, accepted 8 February 2018, ~ doi: 10.1016/j.molap.2018.02.001}

\begin{document}

\begin{frontmatter}

\title{Transmission Spectroscopy with the ACE-FTS Infrared Spectral Atlas of Earth: \\ A Model Validation and Feasibility Study}

\author[imf]{Franz Schreier\corref{ca}}
\ead{franz.schreier@dlr.de}
\author[imf]{Steffen St\"adt}
\author[imf]{Pascal Hedelt}
\author[tub]{Mareike Godolt}
\address[imf]{DLR --- Deutsches Zentrum f\"ur Luft- und Raumfahrt, \\ Institut f\"ur Methodik der Fernerkundung, \\
              Oberpfaffenhofen, 82234 We\ss ling, Germany}
\address[tub]{TUB --- Technische Universit\"at Berlin,
              Zentrum f\"ur Astronomie und Astrophysik,
              Hardenbergstr.\ 36, 10623 Berlin, Germany}
\cortext[ca]{Corresponding author}

\date{February 8, 2018}

\begin{abstract}
Infrared solar occultation measurements are well established for remote sensing of Earth's atmosphere, and the corresponding primary transit spectroscopy has turned out to be valuable for characterization of extrasolar planets.
Our objective is an assessment of the detectability of molecular signatures in Earth's transit spectra.

To this end, we take a limb sequence of representative cloud-free transmission spectra recorded by the space-borne ACE-FTS Earth observation mission (Hughes et al., ACE infrared spectral atlases of
the Earth's atmosphere, JQSRT 2014) and combine these spectra to the effective height of the atmosphere.  These data are compared to spectra modeled with an atmospheric radiative
transfer line-by-line infrared code to study the impact of individual molecules, spectral resolution, the choice of auxiliary data, and numerical approximations.
Moreover, the study serves as a validation of our infrared radiative transfer code.

The largest impact is due to water, carbon dioxide, ozone, methane, nitrous oxide, nitrogen, nitric acid, oxygen, and some chlorofluorocarbons (CFC11 and CFC12).
The effect of further molecules considered in the modeling is either marginal or absent. 
The best matching model has a mean residuum of $0.4\rm\,km$ and a maximum difference of $2\rm\,km$ to the measured effective height.
For a quantitative estimate of visibility and detectability we consider the maximum change of the residual spectrum, the relative change of the residual norm, the additional transit depth, and
signal-to-noise ratios for a JWST setup.
In conclusion, our study provides a list of molecules that are relevant for modeling transmission spectra of Earth-like exoplanets and discusses the feasibility of retrieval.

\end{abstract}



\end{frontmatter}


\section{Introduction}
\label{sec:intro}

More than 20 years after the discovery of the first extrasolar planet around a solar-like star \citep{Mayor95} about 3700 exoplanets have been detected (\url{http://exoplanet.eu/}), including a few dozen super-Earths \citep[e.g.][]{Bean10,Dittmann17}, a few potentially Earth-sized planets (e.g.\ in the Trappist-1 system \citep{deWit16,Gillon17t}, Kepler-22b \citep{Borucki12k}, or K2-137b \citep{Smith18}) and a few Earth-mass planets \citep[e.g.][]{AngladaEscude16,Bonfils18}.
In the last decade the characterization of these remote worlds has attracted increasingly more attention.
The question of the spectral appearance of terrestrial exoplanets and the possibility to identify signatures of life has been the focus of a series of modeling studies, whereas the quantitative characterization by atmospheric retrieval techniques is so far mainly confined to larger objects such as hot Jupiters and Neptune-sized planets.

One of the first comprehensive modeling studies of biosignatures of terrestrial exoplanets has been performed by \citet{DesMarais02} who used a line-by-line (lbl) atmospheric radiative transfer code to assess spectral signatures known from Earth, Mars, and Venus in the mid / thermal infrared (MIR, TIR) and visible to near IR (NIR).
\citet{Segura05} generated synthetic Vis-NIR and TIR spectra of an Earth-like planet orbiting around M dwarfs, and transit spectra of Earth and an Earth-like exoplanet orbiting an M star have been modeled by \citet{Kaltenegger09}.
The impact of the host star's type or distance on the spectral appearance of Earth-like planets has been considered by, e.g., \citet{Segura03,Rauer11,Hedelt13,Vasquez13m,Vasquez13c,Rugheimer15} and \citep{Rugheimer13}.

For an assessment of exoplanet atmospheric remote sensing Earth seen from space is an ideal test case;
in fact it is the only planet that can be used for validation of exoplanet retrieval codes (to some extent, solar system planets, esp.\ Mars and Venus, can be used, too).
\citet{Christensen97} presented observed spectra of Earth obtained with the Thermal Emission Spectrometer (TES) aboard NASA's Mars Global Surveyor at a distance of 4.7 million km.
\citet{Tinetti06m} has compared visible and IR spectra modeled with an lbl multiple scattering code and observations made by TES and the OMEGA instrument aboard ESA's Mars Express as well as ground-based
Earth-shine observations.
Visible or near IR Earth-shine spectra have been analysed by, e.g., \citet{Woolf02, Palle09, GarciaMunoz12e}.
Water vapor and biogenic oxygen and ozone have been identified in high resolution Vis-NIR transmission spectra of Earth seen during the December 2010 Moon eclipse by \citet{Arnold14}.
EPOXI (Extrasolar Planet Observation and Characterization --- Deep Impact eXtended Investigation) observations of Earth have been used by \citet{Kaltenegger07,Rugheimer13,Robinson11} for radiative transfer code validation.
Furthermore, \citet{Irwin14} calculated transit spectra of Earth using the NEMESIS code \citep{Irwin08} and used Rosetta/VIRTIS observations for validation.

To our knowledge, data from space-borne missions dedicated to Earth observation have been rarely used to demonstrate the capabilities of exoplanet atmospheric studies.
For validation of their radiative transfer code, \citet{Misra14} used solar occultation
spectra observed by the ATMOS Fourier transform spectrometer \citep{Abrams96a} on-board the ATLAS 3 Space Shuttle (November 1994) in their assessment of the impact of refraction on transit spectroscopy of Earth-like exoplanets.
Likewise, ATMOS data were used by \citet{Kaltenegger09} for validation purposes.
Disk-averaged MIR spectra have been generated from nadir observations with the Atmospheric Infrared Sounder (AIRS) instrument \citep{Chahine06} aboard NASA's Aqua satellite by \citet{Hearty09},
these data were also used in the \citet{Robinson11} validation study.
\citet{GomezLeal12} studied the disk-integrated MIR thermal emission of Earth seen as a point source using a variety of satellite measurements acquired over more than 20 years.
Solar occultation data obtained with SCIAMACHY (Scanning Imaging Absorption Spectrometer for Atmospheric CHartographY) on ESA's Envisat satellite were considered in \citet{GarciaMunoz12e}.

NIR observations of Venus by SCIAMACHY were used for validation of a line-by-line single scattering code by \citet{Vasquez13v}.
The Swedish small satellite mission Odin originally had a dual role: aeronomy and astronomy; among others, it has been used for measurements of water on Mars and Jupiter \citep{Biver05,Cavalie12}.

Similar to the ATMOS instrument, the Canadian Atmospheric Chemistry Experiment --- Fourier Transform Spectrometer (ACE-FTS) observes the Earth's limb in solar occultation \citep{Bernath05,Bernath17}.
Hundreds of spectra recorded in the 2004 to 2008 time frame have been averaged by \citet{Hughes14} to compile five ``infrared spectral atlases'' for various seasons and latitude bands.

In this study we use the ACE-FTS infrared atlas to generate effective height spectra of Earth's atmosphere and to compare these with model spectra generated with an lbl code in order to assess the
visibility and detectability of atmospheric gases in transit spectra.
In the next section we briefly review the data and models used for this study, and in section \ref{sec:results} we present the results.
First, we study the impact of individual molecular absorbers, auxiliary data, and numerical approximations using the ACE-FTS arctic winter atlas degraded to moderate resolution,
and in subsection \ref{ssec:global} we investigate the molecular visibility and detectability for various resolutions using a mix of all five atlases to mimic a global view.
After a discussion in section \ref{sec:discussion} we summarize our study in section \ref{sec:conclusions} and conclude with a brief outlook.


\section{Data and Models}
\label{sec:theory}

\subsection{ACE-FTS}
\label{ssec:acefts}

The Atmospheric Chemistry Experiment --- Fourier Transform Spectrometer is one of two science instruments aboard the Canadian SCISAT orbiting Earth at $650\rm\,km$ altitude with an inclination of $74^\circ$ \citep{Bernath05,Bernath17}.
ACE-FTS observes the atmospheric absorption in the infrared (750 to $4400 \cm$ or 2.2 to $13.3\mue$) with high spectral resolution ($0.02 \cm$, i.e.\ about two times the ATMOS resolution) in limb viewing geometry.
Each measurement essentially comprises a sequence of transmission spectra for different tangent altitudes and is used to derive altitude dependent profiles of temperature and trace gas concentrations.

\begin{figure*}
 \includegraphics[width=\textwidth]{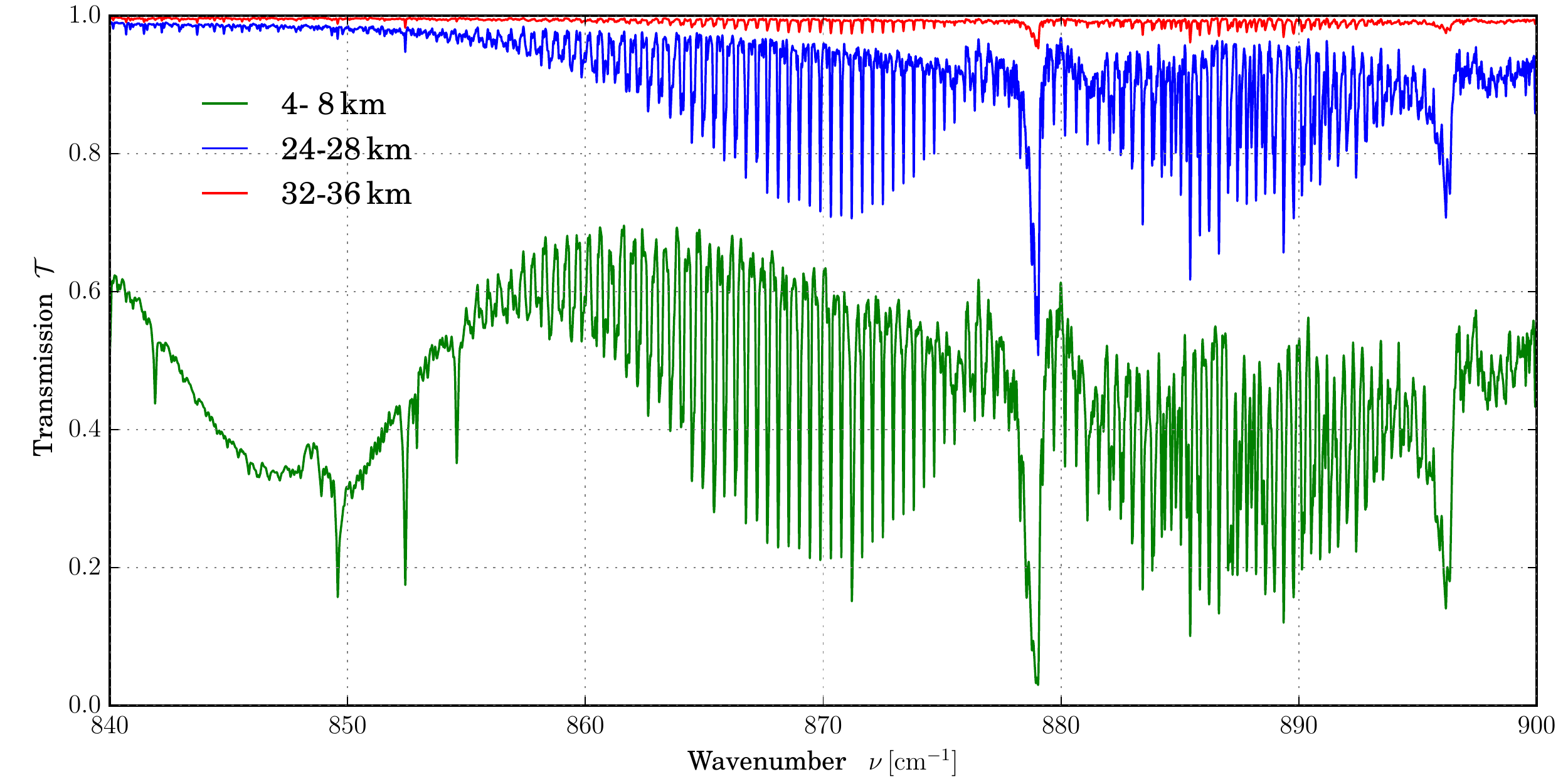}
 \caption{Zoom into three of 31 high resolution limb transmission spectra of the arctic winter atlas. (The large peaks around 879 and $896\cm$ are due to \chem{HNO_3}, the three peaks around $850\cm$ are due to \chem{H_2O}.)}
 \label{fig:zoomArcticWinterTrans}
\end{figure*}

Since its launch in August 2003 ACE-FTS has recorded tens of thousands of occultation spectra.
Representative spectral ``atlases'' of these data have been generated by averaging about 800 cloud-free spectra for a series of altitudes from 4 to $128\rm\,km$ in $4\rm\,km$ steps \citep{Hughes14}.
Five atlases with 31 spectra each are provided for Arctic summer and winter (ASU, AWI; latitude range $60^\circ \TO 90^\circ \rm N$),
midlatitude summer and winter (MLS, MLW; $30^\circ \TO 60^\circ \rm N$), and the tropics ($30^\circ \rm S \TO 30^\circ \rm N$).
The data (with 39\,MB per file/spectrum for 1\,960\,001 wavenumbers in $100 \TO 5000\cm$) are available at \url{http://www.ace.uwaterloo.ca/atlas.php} along with representative plots for 10\,km, 30\,km, 70\,km, and 110\,km tangent heights in pdf format.
\qufig{fig:zoomArcticWinterTrans} shows an excerpt of the arctic winter atlas indicating the high resolution and low noise level (due to the averaging).

Except for a few dips a single ACE-FTS occultation spectrum has a typical signal-to-noise (SNR) ratio of about 100 to 450 for wavenumbers $800 \TO 3500 \cm$ \citep[][Fig.\ 1, $\rm SNR>50$ up to $4000\cm$]{Hughes14}.
Because typically 800 spectra have been averaged, the atlas spectra can be expected to have an SNR of some thousands.

ACE-FTS is characterized by a very high resolution, and the spectra are given with a wavenumber grid spacing $\delta\nu = 0.0025\cm$.
To study the impact of lower resolution, we considered degraded spectra obtained by convolution with a Gaussian response function.  
Initially we take an optimistic point of view and consider spectra convolved with a Gaussian of HWHM (half width at half maximum) $\Gamma=1.0\cm$.
Additionally, we examine the impact of spectral resolution with more realistic HWHMs $\Gamma=2, ~5, ~10, ~20$ and $50\cm$.

\subsection{Infrared radiative transfer}
\label{ssec:irrt}
In a gaseous, non-scattering atmosphere the attenuation of radiation along the path $s$ is described by Beer's law \citep{Zdunkowski07} for transmission $\T$ and optical depth $\tau$ as a function of wavenumber $\nu$,
\begin{align} \label{beer}
\mathcal{T}\left(\nu ,s\right) ~&=~ \E^{-\tau(\nu ,s)} \\
   ~&=~ \exp{\left( -\int\limits_0^s \D s' \sum_m \, k_m\left( \nu,p(s'),T(s') \right) \: n_m(s') \right)} ~, \notag
\end{align}
where $p$ and $T$ are the atmospheric pressure and temperature, and the integrand constitutes the absorption coefficient essentially determined by the sum of the absorption cross sections $k_m$ scaled by the molecular number densities $n_m$.
In high resolution lbl models, the absorption cross section of molecule $m$ is given by the superposition of many lines $l$ with line center positions $\hat\nu_l$, each described by the product
of a temperature--dependent line strength $S_l$ and a normalized line shape function $g$ describing the broadening mechanisms (for brevity the subscript $m$ is omitted),
\begin{equation}  \label{absXS}
 k(\nu,p,T) ~=~ \sum\limits_l S_l(T) \: g\left(\nu; \, \hat\nu_l, \gamma_l(p,T) \right) ~.
\end{equation}
The combined effect of pressure broadening (corresponding to a Lorentz\-ian line shape)  and Doppler broadening (corresponding to a Gaussian line shape) can be represented by a
convolution, i.e.\ the Voigt line profile \citep{Armstrong67,Schreier11v} with width $\gamma\bigl(\gamma_\text{L}(p,T),\gamma_\text{G}(T)\bigr)$.

\subsection{GARLIC}
\label{ssec:garlic}

The ``Generic Atmospheric Radiation Line-by-line Infrared-microwave Code'' \citep{Schreier14,Schreier15} has been developed in the last two decades with an emphasis on efficient and reliable numerical algorithms. 
It is suitable for arbitrary observation geometry, instrumental field-of-view (FoV), and spectral response functions (SRF).
More recently, GARLIC (or its Fortran\,77 predecessor) has also been used for a variety of exoplanet studies, e.g.\ \citep{Rauer11,Hedelt13,Vasquez13m,Vasquez13c}.

The core of GARLIC's subroutines constitutes the basis of forward models used to implement inversion codes to retrieve atmospheric state parameters \citep[e.g.][]{GimenoGarcia11,Xu16}.
For verification, GARLIC has contributed to several intercomparison studies \citep[e.g.][]{Clarmann02,Melsheimer05,Schreier18agk}.
For validation, modeled spectra have been successfully compared with limb thermal emission spectra observed by the MIPAS instrument aboard ENVISAT \citep{Mendrok07} and with Venus observations
\citep{Hedelt11v,Vasquez13v}.

For the lbl computation of cross sections \eqref{absXS}, GARLIC uses a (default) cut-off $\delta\nu=10\cm$ in the line wings ($25\cm$ for \chem{H_2O} with continuum).
Wavenumber grid points are sampled with a (default) spacing of $\delta\nu = \gamma/4$ where $\gamma$ is the (Voigt) half width depending on molecule, pressure, and temperature.
All spectral lines listed in the database are considered, i.e.\ weak lines are not neglected.

\subsection{Spectroscopic data}
\label{sec:specData}

Line parameter databases are a mandatory input for lbl modeling.
For the $700 \TO 4400 \cm$ wavenumber range considered here, the HITRAN 2016 database \citep{Gordon17etal} lists 2.94 million lines of 43  molecules (the entire database has more than nine million lines of 49 molecules),
whereas the GEISA 2015 database \citep{JacquinetHusson16etal} has 2.48 million lines of 51 molecules (total 52 molecules with five million lines). 
Data of molecular masses and rotational and vibrational partition sums (required for the temperature conversion of line strengths) are taken from the ATMOS data set \citep{Norton91}.

In addition to the lbl cross sections described by \qeq{absXS} there is a further contribution commonly known as the continuum \citep{Shine12}.
The pressure and temperature dependent continuum varies slowly with wavenumber and is especially important for water.
Until recently, the continuum implemented in GARLIC was based on code and data extracted from the FASCODE3 lbl model \citep{Clough88} and essentially corresponds to the ``CKD continuum'' \citep{Clough89}
comprising water (self and foreign), carbon dioxide, oxygen and nitrogen contributions.
For the ongoing intercomparison of GARLIC with the ARTS and KOPRA lbl models \citep{Schreier13agk} and for the study reported here an upgrade of the continuum has been made, i.e.\ the
``MT-CKD continuum'' \citep{Mlawer12} (version 2.5) has been implemented.

\subsection{Atmospheric data}
\label{ssec:atmData}

Pressure, temperature and volume mixing ratio profiles for the seven ``main'' gases (corresponding to the molecules considered in the very first version of the HITRAN database \citep{McClatchey73}) and six ``scenarios'' (tropical, midlatitude summer and winter (MLS, MLW), subarctic summer and winter (SAS, SAW), US Standard) have been compiled by \citet{Anderson86} for the $0 \TO 120\rm\,km$ altitude range.
Furthermore, this dataset (a.k.a.\ the ``AFGL profiles'')  provides concentration profiles of further 21 trace gases.

Alternatively, the Committee on Space Research (COSPAR) International Reference Atmosphere (CIRA) provides monthly mean profiles of pressure vs.\ temperature for the altitude range $0 \TO 120\rm\,km$ and latitudes $80^\circ$N to $80^\circ$S with $5^\circ$ spacing \citep[][\url{http://badc.nerc.ac.uk/data/cira/}]{Fleming90}.

Concentration profiles of heavy species, in particular chlorofluorocarbons (CFC) etc., were taken from the MIPAS model atmospheres \citep[][and \url{http://eodg.atm.ox.ac.uk/RFM/atm/}]{Remedios07}.


\subsection{Effective height and transit depth}
\label{ssec:effHeight}

For an exoplanet seen from afar it is not possible to distinguish between limb spectra corresponding to individual tangent heights.
Using (primary) transit observations one essentially measures the effective height
\begin{equation} \label{effHgt}
       h(\nu) ~=~ \int_0^\infty \Bigl( 1 - \T(\nu,z_\text{t}) \Bigr) \, \D z_\text{t}
\end{equation}
where the integral includes all limb transmission spectra with tangent altitude $z_\text{t}$ terminating at ``top-of-atmosphere'' (ToA).
In this study, the ToA of $120\rm\,km$ was defined by the available atmospheric data \citep{Anderson86,Fleming90,Remedios07}.

Numerically, effective height spectra are summed according to the trapezoid quadrature approximation of \eqref{effHgt}
\begin{equation} \label{effHgtSum}
 h(\nu) ~\approx~ \delta z_\text{t} \left( \sum_{l=1}^L \bigl(1-\T(\nu,z_l)\bigr) + {\T(\nu,z_1) + \T(\nu,z_L) \over 2} -1 \right)
\end{equation}%
taken a limb sequence with equidistant tangent points from $z_1$ to $z_L$.
As an alternative to the trapezoid quadrature rule, the integral \eqref{effHgt} can be approximated using a midpoint quadrature.

\begin{figure*}
 \includegraphics[width=\textwidth]{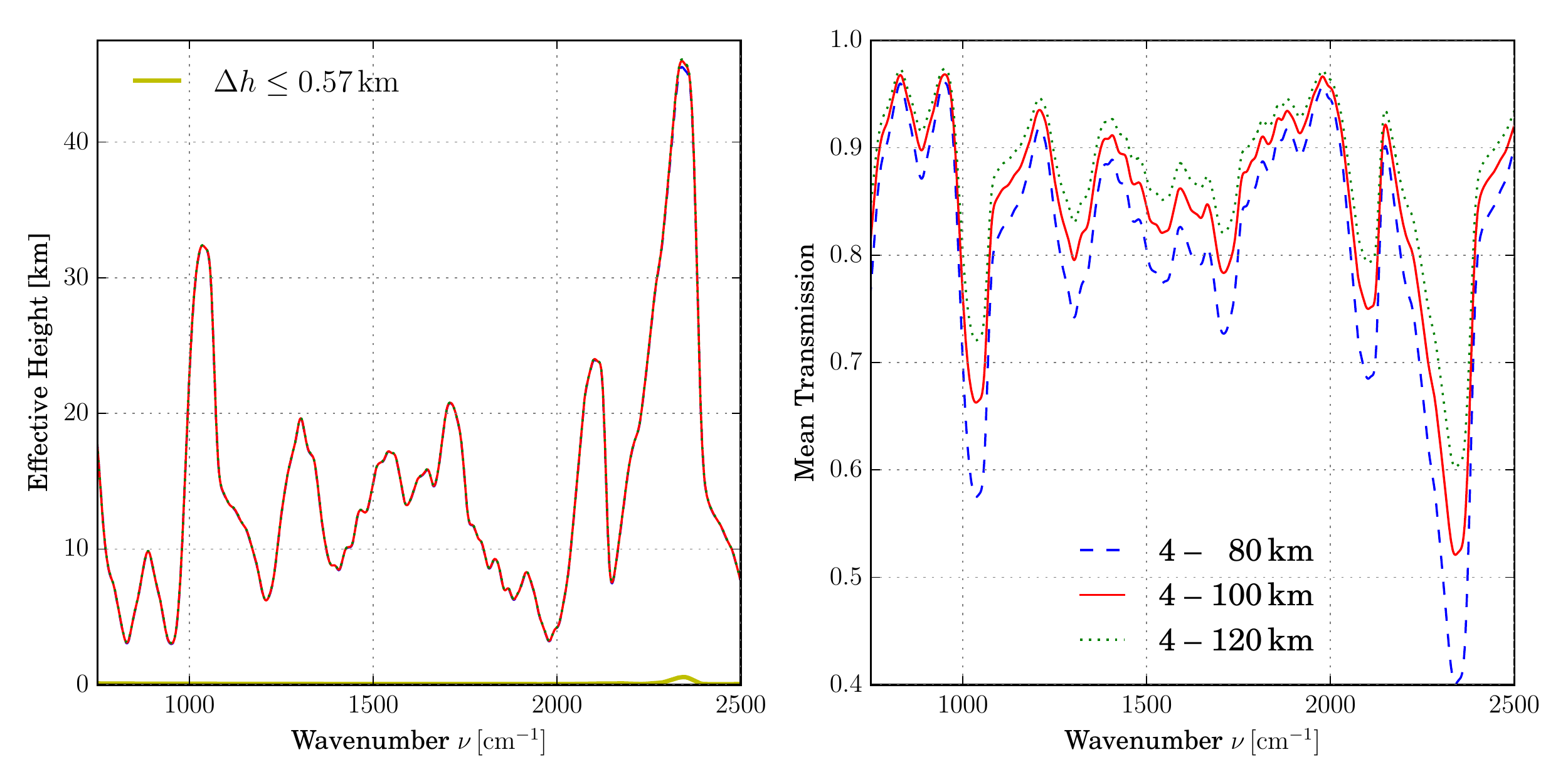}
 \caption{Comparison of ACE-FTS effective height (left) and mean transmission (right) spectra for three different ToA altitudes (arctic winter, smoothed by convolution with a Gaussian of width
$\Gamma=10\cm$). The yellow curve (left plot) is the difference of the effective heights for ToA=120\,km and ToA=80\,km.}
 \label{fig:meanTrans}
\end{figure*}

Some studies of exoplanet transit spectroscopy consider the mean transmission of a limb sequence, $\langle \T \rangle(\nu) = {1 \over L} \sum_l \T_l(\nu)$ \citep[e.g.][]{Kaltenegger09,Rauer11,Tabataba16}.
However, it should be noted that this quantity depends on the choice of the ToA altitude:
Limb rays traversing only the upper layers of the atmosphere have a transmission close to one, and these ones considerably contribute to the overall sum (the division by the number of limb spectra $L$ hardly compensates this), whereas those rays with
$\mathcal{A}(\nu,z_\text{t}) = 1 - \T(\nu,z_\text{t}) \approx 0$ do not contribute significantly to the integral (or sum) in \qeq{effHgt}.
For illustration, \qufig{fig:meanTrans} compares effective height and mean transmission spectra generated from the ACE-FTS data with the ToA ranging from  80 to 120\,km:
The maximum difference of the three effective height spectra is less than one kilometer (i.e.\ essentially indistinguishable in the plot), whereas the mean transmission is clearly larger for the spectrum with a ToA at 120\,km.
Note the small difference of the ToA=120\,km and ToA=80\,km effective height spectra in the center of the \chem{CO_2} band indicating that the mesosphere has some non-negligible
absorption.
Furthermore note that the mean transmission also depends on the choice of the tangent altitudes: for example, a non-equidistant set of tangent points with
dozens of points in the troposphere and just a few in the stratosphere and mesosphere will significantly change the mean transmission.

The additional transit depth due to the atmosphere is defined as \citep{Rauer11,Hedelt13}
\begin{equation} \label{transDepth}
     \delta  d_\text{t}(\nu) = {\bigl( R_\text{p}+h(\nu)\bigr)^2 - R_\text{p}^2 \over R_\text{s}^2}
\end{equation}          
with $R_\text{p}$ and $R_\text{s}$ the planetary and solar radius.
The squared ratio of the radii is also known as the geometric transit depth $d_\text{geo} = R_\text{p}^2/ R_\text{s}^2$.

\subsection{Goodness of fit}
\label{ssec:detect}

For a quantitative estimate of the agreement between observed and modeled effective height spectra the mean and maximum (absolute) difference and the norm of the residual are considered,
\begin{align}
    \langle | \Delta h | \rangle  ~&=~ {1 \over m} \sum_{i=1}^m |  h_\text{obs}(\nu_i) - h_\text{mod}(\nu_i) | \label{defMean} \\
            \Delta h_\text{max} ~&=~ \max_i | h_\text{obs}(\nu_i) - h_\text{mod}(\nu_i) | \label{defMax} \\
    \Vert \Delta h\Vert          ~&=~      \left[ \sum_{i=1}^m \bigr(  h_\text{obs}(\nu_i) - h_\text{mod}(\nu_i) \bigl)^2 \right]^{1/2} \label{defNorm} 
\end{align}
where $m$ is the number of data points in the spectra.
Whereas $\langle | \Delta h| \rangle$ and $\Delta h_\text{max}$ are more intuitive, the residual norm $\Vert \Delta h\Vert$ is the quantity to be minimized in a least squares fitting procedure.
Note that the mean residuum is largely independent of the number $m$ of data points, and the maximum residual is linked to the spectral resolution.

The detectability of a certain molecule will depend on the spectral resolution and the noise of the observation.
For a quantitative assessment the relative change of the residual norm due to the inclusion/exclusion of a molecule will be used, i.e.\ 
\begin{equation} \label{relNormchange}
(\Vert \Delta h^{m+1} \Vert - \Vert \Delta h^m \Vert) / \Vert \Delta h^m \Vert
\end{equation}
with $m$ indicating the number of absorbing molecules.
This is in analogy to the ``$y$--convergence'' test considering the relative change of the residuals from iteration to iteration, that is commonly used for the iterative solution of nonlinear least squares problems \citep{Dennis81a,Boggs89}:
if this ratio is smaller than a given threshold $\epsilon$ (essentially proportional to the reciprocal of the SNR), the iteration is stopped.

\subsection{Signal-to-Noise}
For an estimate of the signal-to-noise ratio to be expected for Earth seen from afar, we use the noise model of \citet{Rauer11}.
The $\text{SNR}_\text{T}$ of a spectral feature observed in transmission is the product of the stellar $\text{SNR}_\text{S}$ and the change of the additional transit depth $\Delta \delta d_\text{t}$,
\begin{equation} \label{snr}
 \text{SNR}_\text{T} = \text{SNR}_\text{S} \, \Delta\delta d_\text{t}
\quad\text{with}\quad
\text{SNR}_\text{S} = \sqrt{{R_\text{S}^2 \over D^2} I_\text{S} A t  \, {\lambda^2 \over hc} {q \over R}} ~,
\end{equation}
where $D$ is the observer-star distance,
$I_\text{S}$ the stellar spectral energy flux,
$A$ the telescope area,
$t$ the integration time,
$R$ the resolving power,
$q$ a measure of the instrument's throughput,
$\lambda$ wavelength,
and $h$, $c$ Planck's constant and the speed of light, respectively.

\begin{figure*}
 \includegraphics[width=\textwidth]{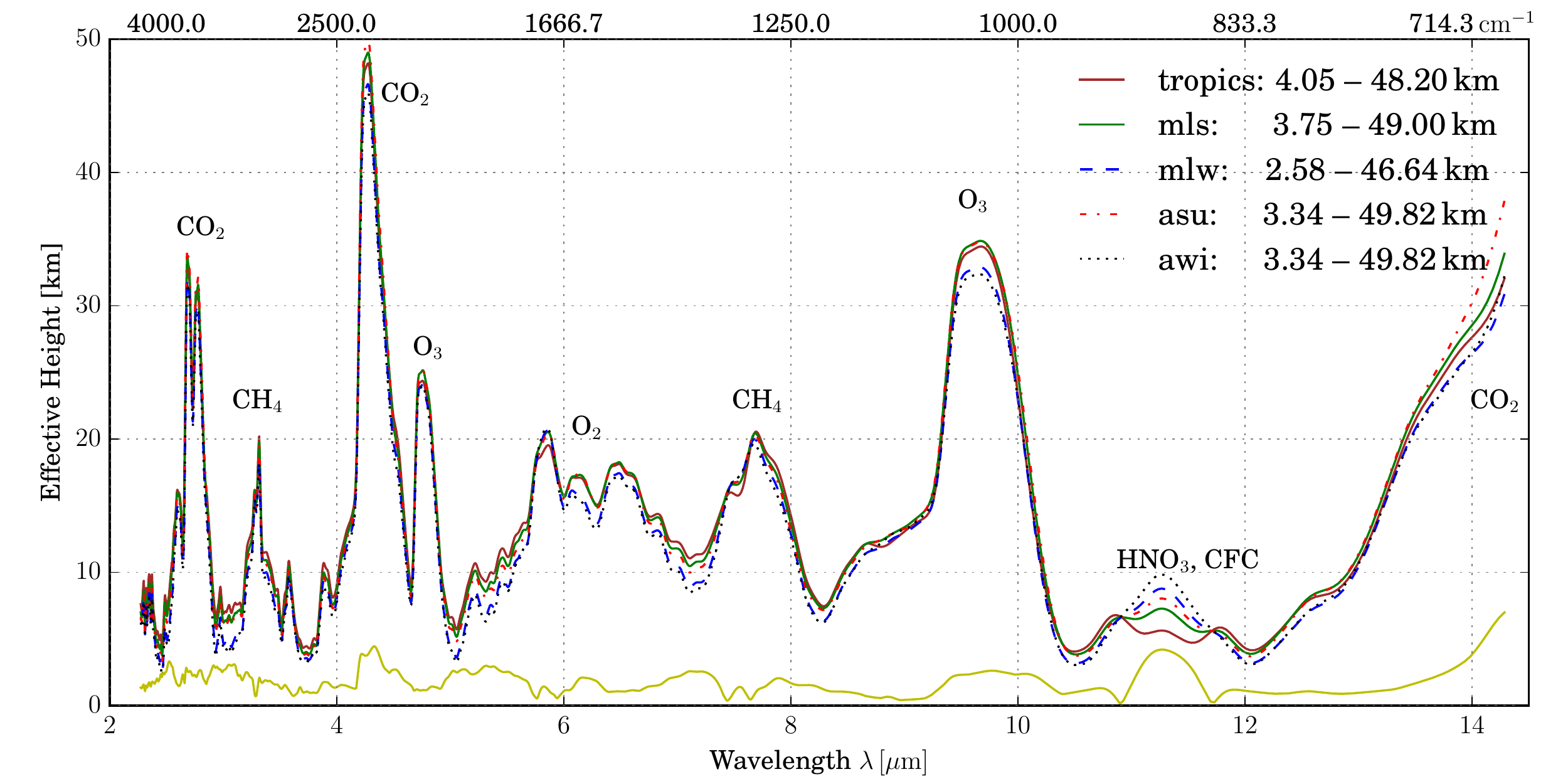}
 \caption{Comparison of effective height spectra resulting from the five IR atlases. The yellow line indicates the maximum difference $h_\text{max} - h_\text{min}$.
          The lower $x$-axis shows wavelengths in micrometer whereas the upper axis shows wavenumbers $\nu = 1/\lambda$ in $\cm$.
          The numbers in the legend indicate the minimum and maximum effective height. Molecules responsible for some of the major features are indicated (\chem{H_2O} absorbs ``almost everywhere'' and is not indicated).}
 \label{fig:effHeightObs}
\end{figure*}


\section{Results}
\label{sec:results}

The observed effective height spectra derived from the five IR atlases by combination of all limb transmission spectra according to \qeq{effHgt} and smoothing with a Gauss of half width $\Gamma=10\cm$ are compared in \qufig{fig:effHeightObs}.
Ignoring the long wavelength end, the difference between the heights can be as large as a few kilometers.
Large differences occur for relative transparent spectral regions as well as for regions characterized by strong absorption (e.g.\ the two \chem{CO_2} bands and the ozone band at $10\mue$).
Also note the large spread of the effective heights in the \chem{HNO_3} band around $11\mue$ with strongly enhanced absorption for the arctic winter. 

\begin{table}
\caption{Comparison of mean, extremum, and norm residual for the moderate resolution arctic winter runs as a function of the number of absorbing molecules.
         The ``relNorm'' column gives the relative change of the norm as defined in \eqref{relNormchange}, all other columns give heights in kilometers.
         Except for the ``GEISA'' column all model runs have been using the HITRAN database.
         A tripled \chem{HNO_3} and doubled CFC11, CFC12 concentrations have been used.
         The very first row is for \chem{CO_2} alone. The second column identifies the molecule added to the list of absorbers.}
\label{tab:residuaGaussAWI1}
\begin{tiny}
\begin{tabular}{clrrrrrrrrr}
\hline
$m$ &          &    mean  & max    &  norm    &    GEISA &   relNorm  & $\Delta h_\text{max}$ \\
\hline
  1 &     ~CO2 &    8.003 & 33.849 &  1219.99 &  1219.65 &            &          \\
  1 &     ~H2O &    7.978 & 47.961 &  1460.76 &  1461.84 &            &          \\
  2 &     +CO2 &    4.721 & 33.616 &   893.33 &   893.68 &     0.3884 &   48.884 \\
  3 &      +O3 &    3.058 & 22.085 &   564.35 &   565.06 &     0.3683 &   34.003 \\
  4 &     +N2O &    2.381 & 19.943 &   445.09 &   446.52 &     0.2113 &   14.363 \\
  5 &      +CO &    2.341 & 19.943 &   442.00 &   443.48 &     0.0069 &    3.523 \\
  6 &     +CH4 &    1.141 & 11.213 &   244.23 &   245.61 &     0.4474 &   21.385 \\
  7 &      +O2 &    1.022 & 11.213 &   230.08 &   231.18 &     0.0580 &    4.813 \\
  8 &      +NO &    1.020 & 11.213 &   230.05 &   231.15 &     0.0001 &    0.355 \\
  9 &     +SO2 &    1.020 & 11.213 &   229.98 &   231.09 &     0.0003 &    0.089 \\
 10 &     +NO2 &    1.014 & 11.212 &   229.74 &   230.81 &     0.0010 &    2.307 \\
 11 &     +NH3 &    1.013 & 11.212 &   229.51 &   230.58 &     0.0010 &    0.226 \\
 12 &    +HNO3 &    0.773 &  6.914 &   161.74 &   163.22 &     0.2953 &   10.381 \\
 13 &      +OH &    0.773 &  6.914 &   161.74 &   163.22 &     0.0000 &    0.008 \\
 14 &      +HF &    0.773 &  6.914 &   161.76 &   163.24 &     0.0001 &    0.236 \\
 15 &     +HCl &    0.772 &  6.914 &   161.70 &   163.17 &     0.0004 &    0.251 \\
 16 &     +HBr &    0.772 &  6.914 &   161.69 &   163.17 &     0.0000 &    0.003 \\
 17 &      +HI &    0.772 &  6.914 &   161.69 &   163.17 &     0.0000 &    0.002 \\
 18 &     +ClO &    0.772 &  6.914 &   161.69 &   163.17 &     0.0000 &    0.001 \\
 19 &     +OCS &    0.764 &  6.914 &   161.06 &   162.61 &     0.0039 &    1.138 \\
 20 &    +H2CO &    0.762 &  6.914 &   161.00 &   162.56 &     0.0003 &    0.058 \\
 21 &    +HOCl &    0.762 &  6.914 &   160.97 &   162.52 &     0.0002 &    0.027 \\
 22 &      +N2 &    0.575 &  5.424 &   108.03 &   110.58 &     0.3289 &    7.984 \\
 23 &     +HCN &    0.573 &  5.424 &   107.99 &   110.53 &     0.0004 &    0.127 \\
 24 &   +CH3Cl &    0.572 &  5.424 &   107.86 &   110.40 &     0.0012 &    0.125 \\
 25 &    +H2O2 &    0.572 &  5.424 &   107.81 &   110.34 &     0.0005 &    0.031 \\
 26 &    +C2H2 &    0.572 &  5.424 &   107.81 &   110.34 &     0.0001 &    0.013 \\
 27 &    +C2H6 &    0.569 &  5.424 &   107.23 &   109.59 &     0.0053 &    0.378 \\
 28 &     +PH3 &    0.569 &  5.424 &   107.23 &   109.59 &     0.0000 &    0.000 \\
 29 &    +COF2 &    0.566 &  5.424 &   106.89 &   109.25 &     0.0032 &    0.142 \\
 30 &     +SF6 &    0.566 &  5.424 &   106.82 &   109.22 &     0.0007 &    0.174 \\
 31 &   +CCl3F &    0.543 &  5.424 &   100.70 &   103.24 &     0.0573 &    4.919 \\
 32 &  +CCl2F2 &    0.489 &  2.635 &    81.94 &    84.95 &     0.1863 &    5.019 \\
 33 &   +CClF3 &    0.489 &  2.635 &    81.94 &    84.95 &     0.0000 &    0.000 \\
 34 &     +CF4 &    0.487 &  2.635 &    81.67 &    84.56 &     0.0033 &    1.081 \\
 35 & +C2Cl3F3 &    0.485 &  2.630 &    81.14 &    84.04 &     0.0066 &    0.139 \\
 36 &  +CHClF2 &    0.480 &  2.627 &    79.87 &    82.80 &     0.0156 &    0.214 \\
 37 &  +ClONO2 &    0.472 &  2.376 &    77.97 &    80.86 &     0.0237 &    0.841 \\
 38 &    +N2O5 &    0.462 &  2.375 &    76.06 &    78.84 &     0.0246 &    0.969 \\
\hline
\end{tabular}
\end{tiny}
\end{table}

The volume mixing ratio of $330\rm\,ppm$ and $1.7\rm\,ppm$ of carbon dioxide and methane, respectively, specified in the \citet{Anderson86} dataset are clearly outdated for modeling the ACE-FTS
measurements covering the 2004 \TO 2008 time range.
Visual inspection of the observed and modeled effective height spectra and their residuals indicates that $380\rm\,ppm$ for \chem{CO_2} and a \chem{CH_4} profile scaled by a factor 1.3 yield the best agreement.
For recent satellite and ground-based measurements of these carbon species see, e.g.,\citep{Butz11}.

\subsection{The arctic winter atlas}
\label{ssec:arcticWinter}

In the following we will first study the effective height spectrum resulting from the combined arctic winter IR atlas for two reasons:
First, water is highly variable in Earth's atmosphere, and it is hence difficult to select a representative \chem{H_2O} profile.
Furthermore, laboratory spectroscopy of water is not trivial mainly because of the difficulty to exactly quantify the amount of water in the gas absorption cell.
The appropriate line shape, uncertainties due to pressure broadening parameters and due to continuum contributions further complicate matters \citep{Gordon07,Bailey09,Barton17,Yang16}.
Accordingly we have chosen the arctic winter case characterized by a relatively dry atmosphere.
The subarctic winter profile of \citet{Anderson86} has an integrated water content of $4.1 \rm\, kg/m^2$ compared to $40.7 \rm\, kg/m^2$ for the tropical profile.

Unless otherwise noted, the subarctic winter pressure, temperature and concentration profiles of the main absorbers \citep{Anderson86} were used along with spectroscopic data (lbl and cross
sections) from HITRAN\,2016 and CKD continuum data.
\begin{table}[t!]
\caption{Impact of a missing molecule on the residual norm for the moderate resolution arctic winter runs.
         The ``relNorm'' column gives the relative change of the norm with the 38 molecules spectrum as reference (in analogy to \qeq{relNormchange} with $m=37$).
         The third column is the maximum change of the effective height $\delta h = h_{38} - h_{37}$.}
\label{tab:residuaGaussAWI2}
\begin{scriptsize}
\begin{tabular}{lrrr}
\hline
         &    norm    & relNorm & $\delta h_\text{max}$ \\
         &    [km]    &         & [km]       \\
\hline
    none &      76.06 &         &         \\
     H2O &     448.42 &   4.896 &  19.219 \\
     CO2 &     790.42 &   9.392 &  35.288 \\
      O3 &     620.50 &   7.158 &  33.801 \\
     N2O &     156.49 &   1.057 &   7.813 \\
      CO &      76.19 &   0.002 &   2.657 \\
     CH4 &     339.33 &   3.461 &  21.366 \\
      O2 &      95.78 &   0.259 &   4.596 \\
      NO &      76.12 &   0.001 &   0.354 \\
     SO2 &      76.07 &   0.000 &   0.088 \\
     NO2 &      76.63 &   0.007 &   2.307 \\
     NH3 &      76.18 &   0.002 &   0.199 \\
    HNO3 &     165.68 &   1.178 &  10.328 \\
      OH &      76.06 &   0.000 &   0.008 \\
      HF &      76.02 &   0.001 &   0.236 \\
     HCl &      76.18 &   0.002 &   0.250 \\
     HBr &      76.06 &   0.000 &   0.001 \\
      HI &      76.06 &   0.000 &   0.003 \\
     ClO &      76.06 &   0.000 &   0.001 \\
     OCS &      77.06 &   0.013 &   1.137 \\
    H2CO &      76.15 &   0.001 &   0.058 \\
    HOCl &      76.10 &   0.001 &   0.027 \\
      N2 &     141.50 &   0.860 &   7.983 \\
     HCN &      76.10 &   0.001 &   0.127 \\
   CH3Cl &      76.20 &   0.002 &   0.119 \\
    H2O2 &      76.09 &   0.000 &   0.030 \\
    C2H2 &      76.07 &   0.000 &   0.013 \\
    C2H6 &      76.55 &   0.006 &   0.378 \\
     PH3 &      76.06 &   0.000 &   0.000 \\
    COF2 &      76.39 &   0.004 &   0.140 \\
     SF6 &      76.14 &   0.001 &   0.173 \\
   CCl3F &      83.13 &   0.093 &   4.904 \\
  CCl2F2 &      94.96 &   0.248 &   4.976 \\
   CClF3 &      76.06 &   0.000 &   0.000 \\
     CF4 &      76.26 &   0.003 &   1.059 \\
 C2Cl3F3 &      76.58 &   0.007 &   0.138 \\
  CHClF2 &      77.36 &   0.017 &   0.214 \\
  ClONO2 &      77.97 &   0.025 &   0.841 \\
    N2O5 &      77.97 &   0.025 &   0.969 \\
\hline
\end{tabular}
\end{scriptsize}
\end{table}

\subsubsection{Adding molecule by molecule}
\label{sssec:molecules}

A key question of exoplanet science is the detectability of various molecules, esp.\ biosignatures, by atmospheric remote sensing.
Not surprisingly the carbon dioxide bands at $4.3$ and $15\mue$ (partially only), the water band at $6.3 \mue$, and the combined \chem{H_2O} and \chem{CO_2} absorption at $2.7\mue$ are readily
recognizable, see \qufig{fig:effHeightObs}.
Moreover, the atmospheric window around $10\mue$ is visible, albeit with interruptions due to ozone and nitric acid absorption features.

For an assessment of the impact of atmospheric gases on the observed spectrum a series of spectra has been modeled, starting with \chem{H_2O} or \chem{CO_2} alone, and adding molecule-by-molecule in the order of the HITRAN lbl database up to $m\le 30$ (i.e.\ \chem{SF_6}).
Note that pressure and temperature are left unchanged, i.e.\ the surface pressure is about 1\,bar for all runs.
The progress is monitored by the mean, maximum, and norm residuals listed in \qutab{tab:residuaGaussAWI1}.

\begin{figure*}[t]
 \includegraphics[width=\textwidth]{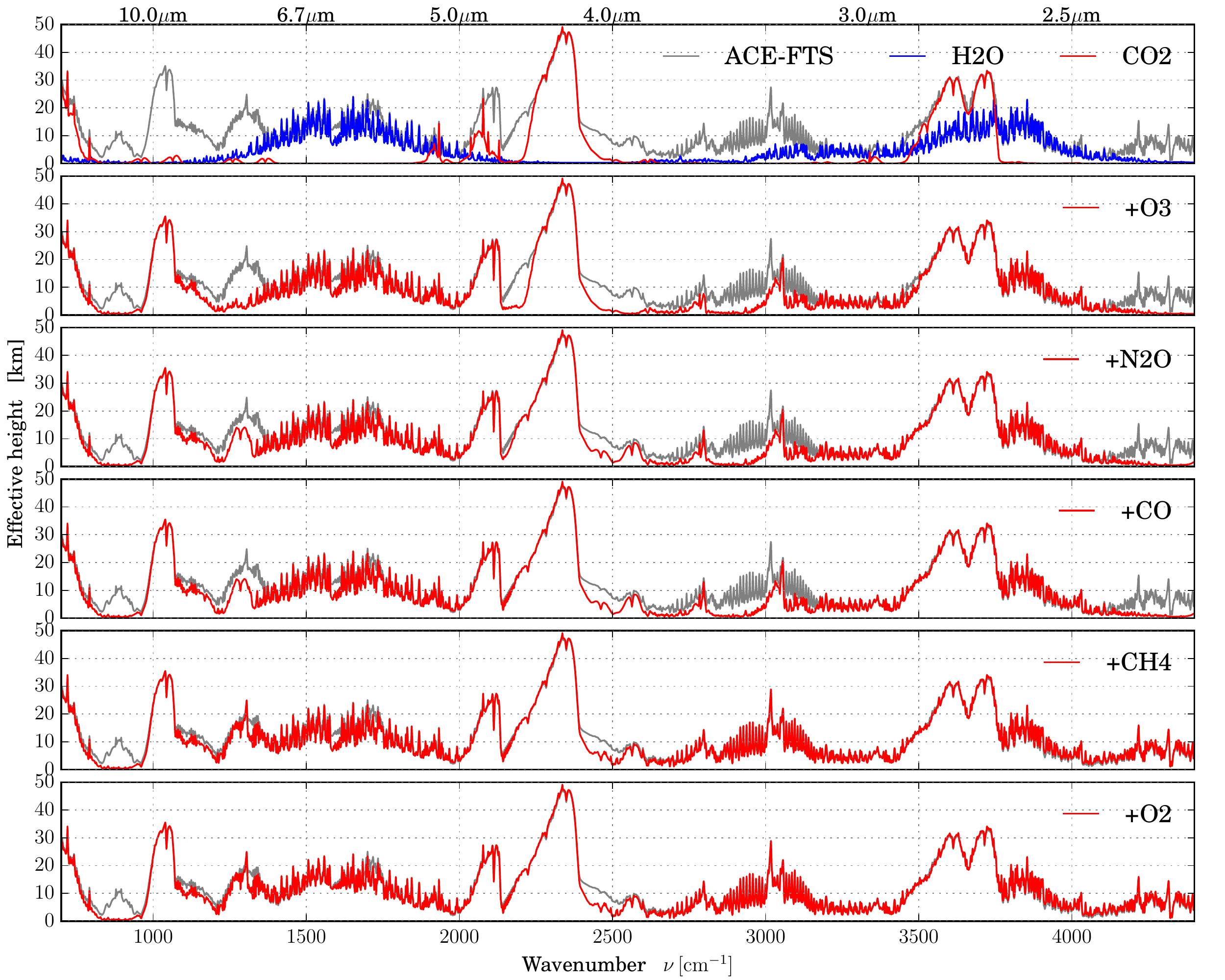}
 \caption{Effective height spectra for the main gases (arctic winter, moderate resolution $\Gamma=1\cm$).
          In the top the model spectra of pure water or carbon dioxide atmospheres are compared to the ACE-FTS observation.
          In the following plots absorption due to the species indicated in the legend is added. The bottom shows the model spectrum for seven absorbers.}
 \label{fig:residua_main}
\end{figure*}

The effective height spectrum modeled with \chem{H_2O}, \chem{CO_2} and \chem{O_3} already shows the main features of the observed effective height spectrum, esp.\ the largest peaks around the
centers of their main absorption bands, cf.\ \qufig{fig:residua_main} (first and second from top).
Note that although water is mainly present in the (lower) troposphere, the pure \chem{H_2O} effective height spectrum reaches up to $20\rm\,km$ in its absorption bands.
Inclusion of nitrous oxide \chem{N_2O} significantly reduces the discrepancies at its fundamental bands around 1250 and $2200\cm$, and less drastically at 2500 and $3500\cm$.
The impact of carbon monoxide is clearly seen for the first fundamental band at $2150\cm$, whereas the first overtone around $4250\cm$ is barely visible,
and the change of the residual mean and norm is minimal.
For methane the largest contribution to the spectrum occurs in the pentad band around $3000\cm$ (from $\Delta h$ as large as $20\rm\,km$ to almost zero), but the reduction in the dyad
($1300\cm$) and octad ($4300\cm$) is also quite dramatic with more than $10 \rm\,km$ in the peaks.
The strong reduction of the discrepancy around $1600\cm$ is due to oxygen (\qufig{fig:residua_main} bottom).

\begin{figure*}
 \centering\includegraphics[width=0.8\textwidth]{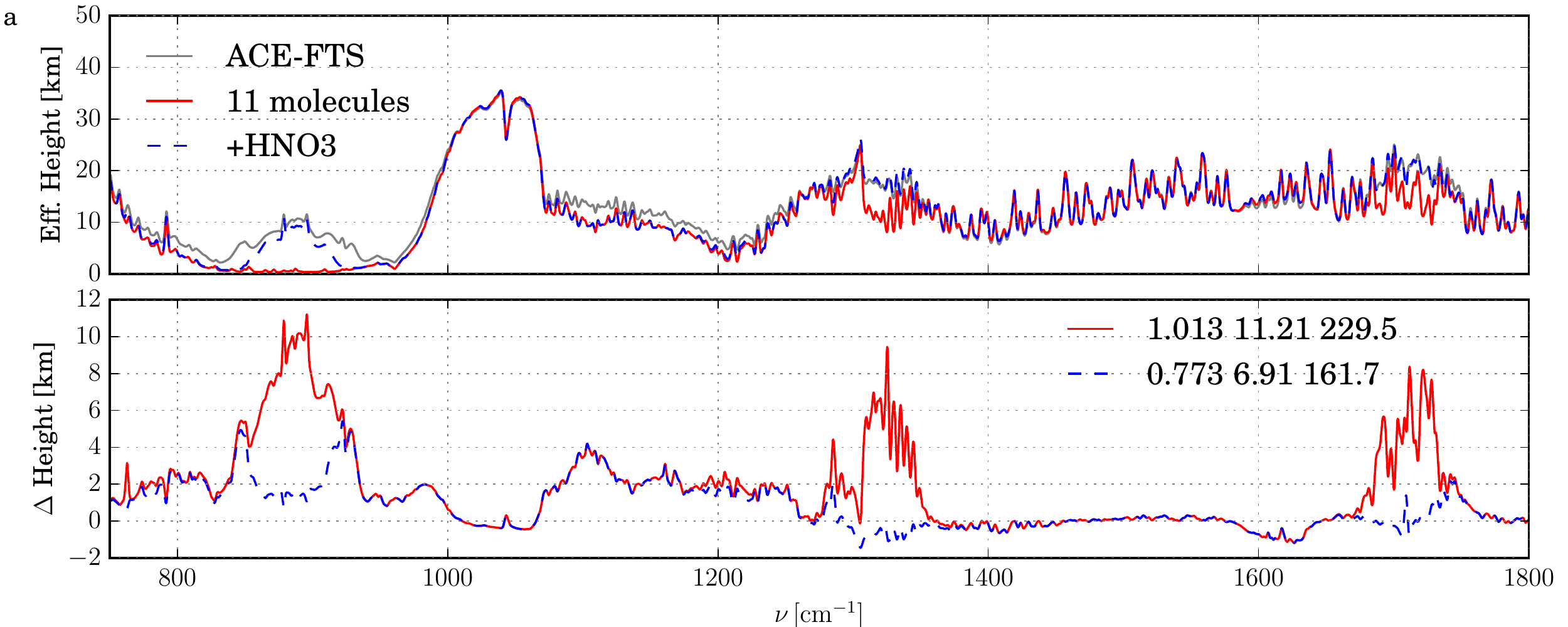}
 \\[-2ex]
 \centering\includegraphics[width=0.8\textwidth]{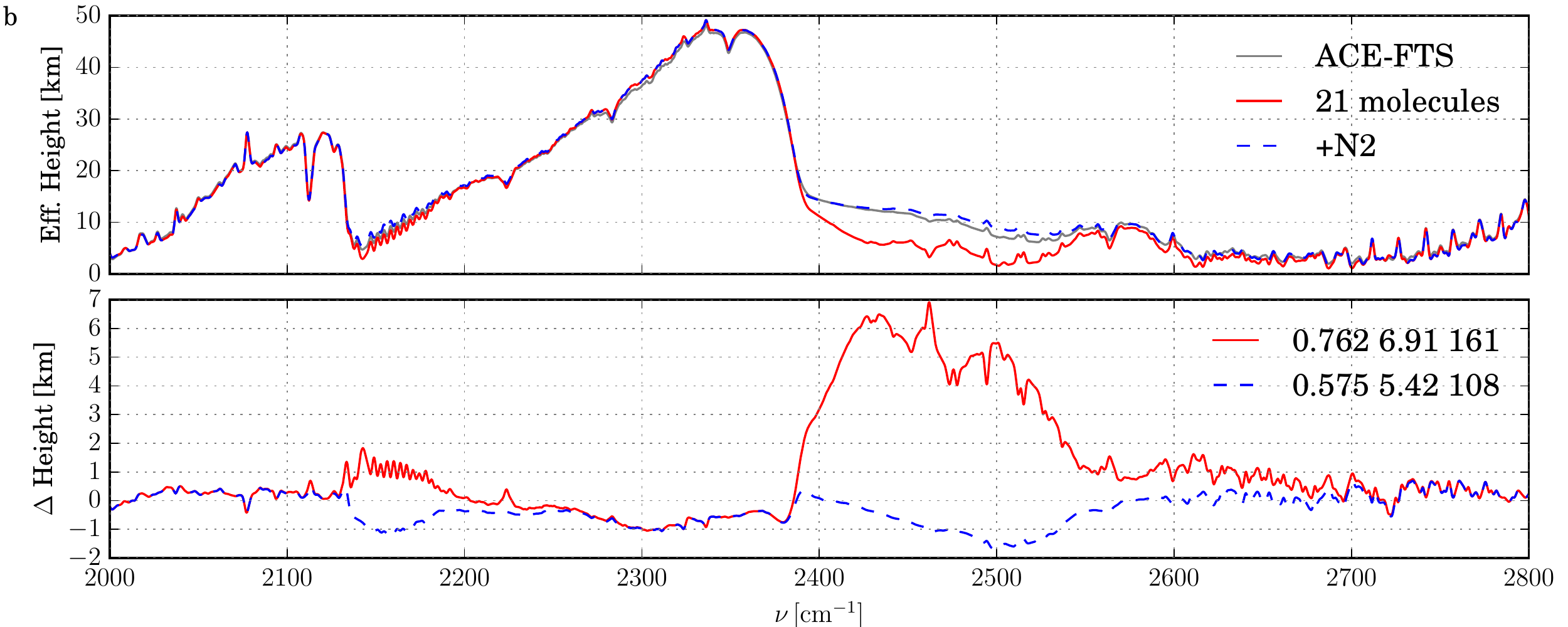}
 \\[-2ex]
 \centering\includegraphics[width=0.8\textwidth]{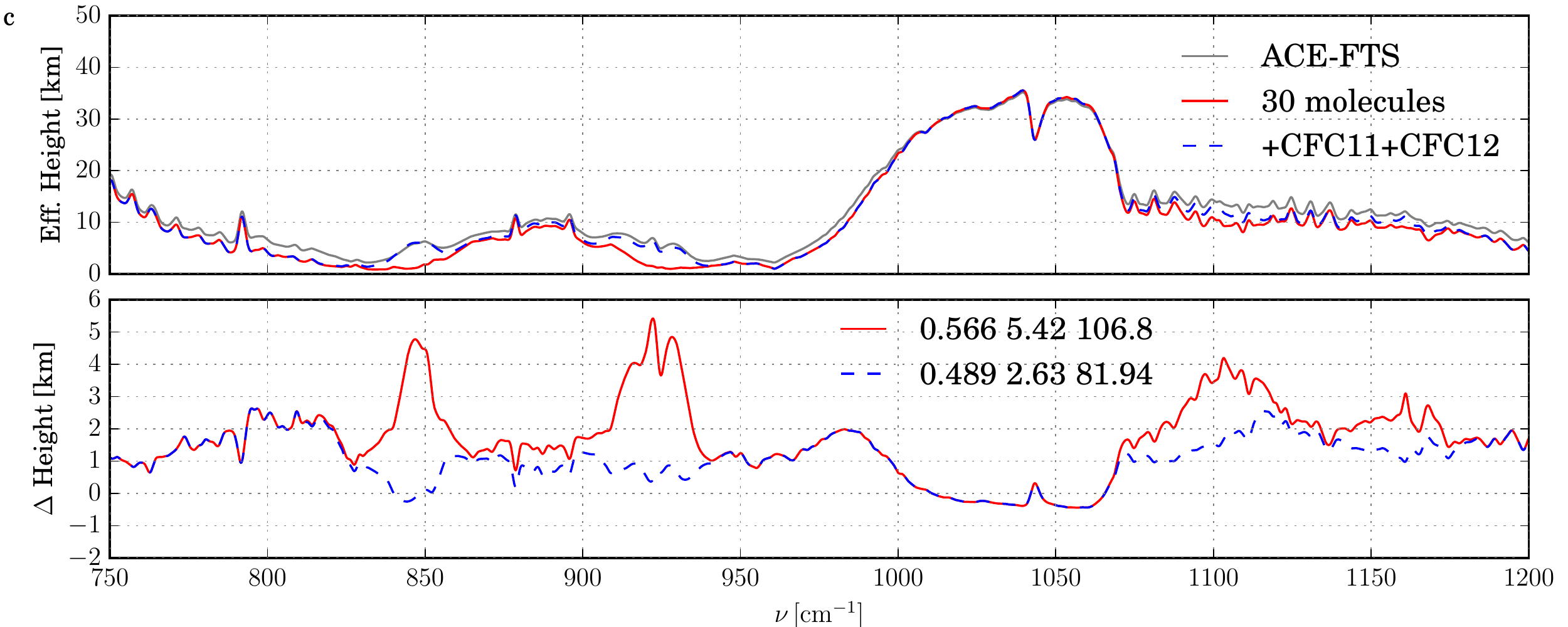}
 \\[-2ex]
 \centering\includegraphics[width=0.8\textwidth]{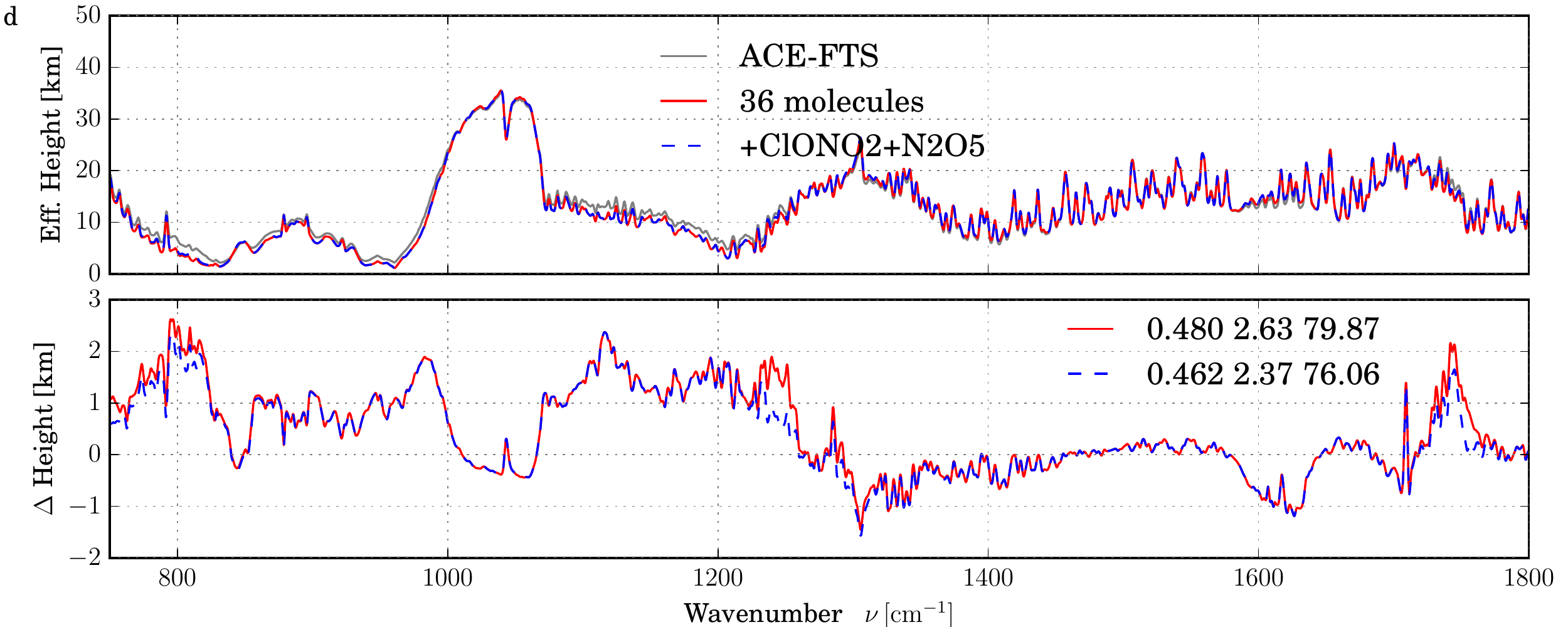}
 \caption{Zoom into effective height spectra and residuals:
          a) impact of nitric acid;
          b) nitrogen;
          c) CFC11 and CFC12;
          d) chlorine nitrate and nitrogen pentoxide.
          The numbers in the legend are the mean \eqref{defMean}, maximum \eqref{defMax}, and norm \eqref{defNorm} residual computed for the whole spectrum.}
 \label{fig:residua_add}
\end{figure*}

The following four molecules in the HITRAN list (\chem{NO}, \chem{SO_2}, \chem{NO_2}, and \chem{NH_3}) do not significantly reduce the residual.
The large underestimate of the effective height around $900\cm$ can be largely attributed to nitric acid, see \qufig{fig:residua_add}a).
The default profile given by \citet{Anderson86} turned out to be inadequate and increasing the \chem{HNO_3} mixing ratio by a factor four leads to a satisfying agreement in this band;
however, this also yields an excess absorption at 1300 and $1700 \cm$, and a tripled nitric acid concentration is used henceforth.
Indeed, the \chem{HNO_3} profiles collected in the ACE-FTS Climatology \citep{Jones12,Koo17} confirm elevated concentrations for Northern latitudes, and the polar summer and winter \chem{HNO_3}
concentrations of the MIPAS model atmospheres \citep{Remedios07} are about a factor two larger than the midlatitude summer and winter profiles (which are identical to the profiles used here \citep{Anderson86}).
Enhanced nitric acid total columns in polar winter have also been measured by the Infrared Atmospheric Sounding Interferometer (IASI) \citep{Wespes09}.
Note that the chlorofluorocarbons CFC11 and CFC12 also have strong absorption in the $830 \TO 930\cm$ interval, see discussion below.

The next large reduction of the residual is due to the nitrogen molecule, where both lbl and continuum contribute to the absorption in the $1900 \TO 2800\cm$ region (\qufig{fig:residua_add}b).
As \qutab{tab:residuaGaussAWI1} indicates, inclusion of lbl contributions from more molecules (23 to 29) does not further reduce the residual remarkably.

To further improve the model spectra it is necessary to consider contributions from molecules with spectroscopic properties available as cross sections (instead of lbl data) only.
Note that for \chem{SF_6} HITRAN line data are provided in a supplementary folder only, and the use of cross section data is recommended.
The CFCs \chem{CCl_3F} and \chem{CCl_2F_2} both have strong bands in the region of the \chem{HNO_3} band mentioned above:
Adding CFC11 (\chem{CCl_3F}) reduces the residual norm from 106.8 to $100.7\rm\,km$, but leaves the largest discrepancy of $\Delta h_\text{max} \approx 5.2\rm\,km$ at about $922\cm$ unchanged;
CFC12 (\chem{CCl_2F_2}) has its peak absorption here, and its inclusion largely eliminates this deviation and further reduces the residual norm to $82\rm\,km$, compare \qufig{fig:residua_add}c. 

The contributions of further heavy molecules, in particular \chem{N_2O_5}, have also a small impact on the effective height spectrum.
The final mean and norm residual (\qufig{fig:residua_add}d) for 38 molecules is $0.46\rm\,km$ and $76\rm\,km$, respectively, compared to 1.02 and $230\rm\,km$ for the first seven HITRAN (or GEISA) molecules.

Adding the CFCs to the model atmosphere actually compensates for some of the extra \chem{HNO_3} required to reduce the discrepancies around $900\cm$.
Scaling both the CFC11 and CFC12 profile by a factor two and scaling \chem{HNO_3} by a factor three gives the smallest residuum with $0.46\rm\,km$ mean residuum and norm $76.1\rm\,km$.
Note that enhanced CFC concentrations are confirmed by the polar winter profiles given in the MIPAS climatology \citep{Remedios07}.

\subsubsection{Spectroscopic input data}
\label{sssec:specData}

\begin{figure*}[t]
 \centering\includegraphics[width=0.9\textwidth]{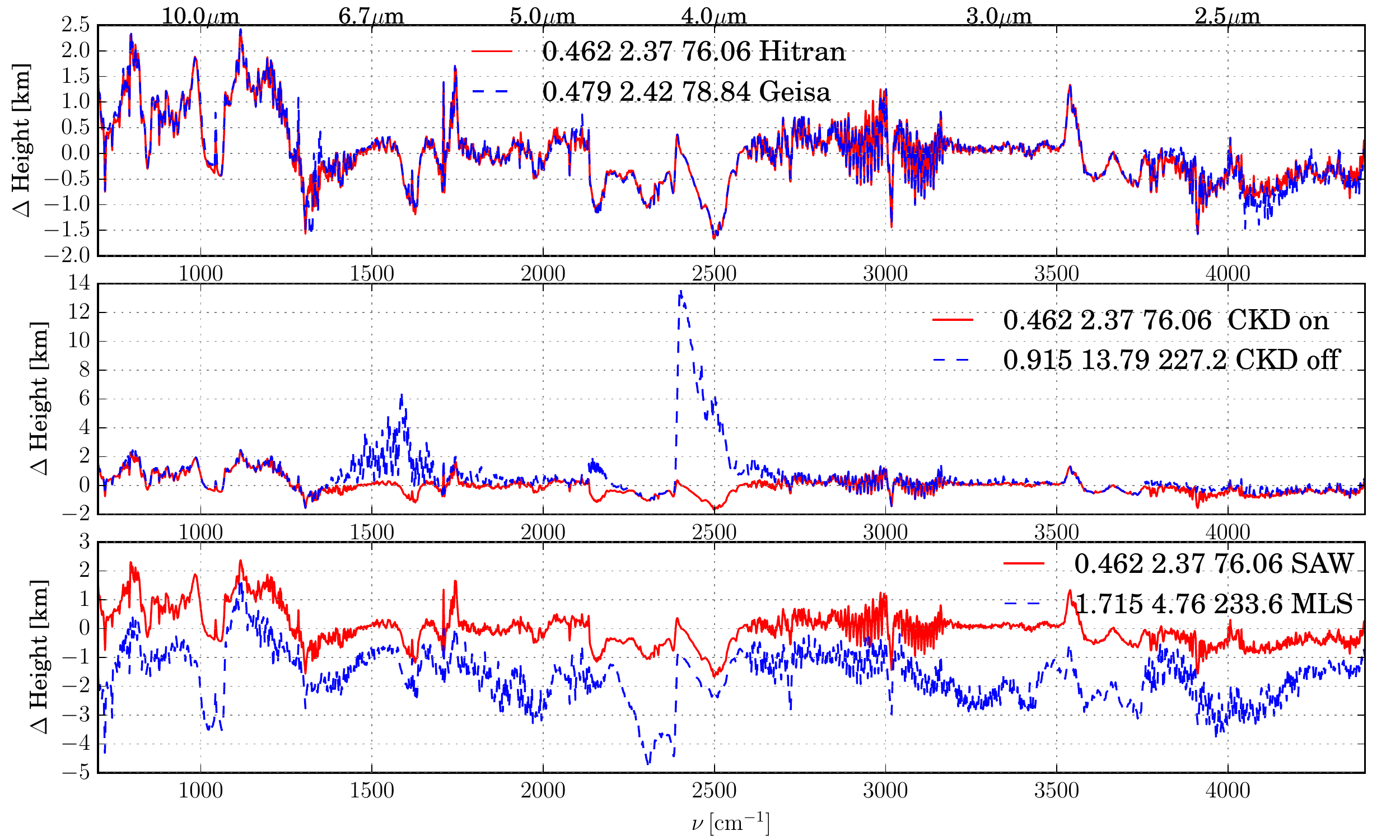}
 \caption{Impact of auxiliary data on effective height residuals for 38 molecules \chem{H_2O}, \dots, \chem{N_2O_5}:
          Top: HITRAN (red) vs.\ GEISA (blue);
          Mid: CKD continuum on (red) vs.\ off (blue);
          Bottom: SAW (red) vs.\ MLS (blue) pressure, temperature, and concentrations.
          Legend numbers as in \qufig{fig:residua_add}.}
 \label{fig:auxData}
\end{figure*}

The GEISA database \citep{JacquinetHusson16etal} is a widely used alternative to HITRAN in Earth science.
The effective height residuals shown in \qufig{fig:auxData} (top) indicate that line parameters of HITRAN\,16 yield a slightly better agreement with the observations compared to GEISA\,15.
\qutab{tab:residuaGaussAWI1} and some model runs with input data mixed from both databases show that this superiority can be largely attributed to the different \chem{H_2O} spectroscopic data, for all the other molecules both databases perform equally well.
This is also confirmed by a comparison of model spectra with 37 absorbing molecules, i.e.\ all molecules except for water (see the subsection \ref{sssec:topDown} below), giving almost identical residuals.
Note that with HITRAN\,12 (with 2.3 million lines contributing) the residuals are slightly smaller, with a mean and norm 0.455 and $75.62\rm\,km$, respectively.
Also note that, in contrast to HITRAN, the GEISA database contains 46031 \chem{SF_6} lines in $920 \TO 976\cm$ (the HITRAN supplementary listing comprises almost three million \chem{SF_6} lines).

Inclusion of continuum absorption in the radiative transfer modeling is clearly important, as demonstrated by the comparison of effective height spectra modeled with and without continuum in \qufig{fig:auxData} (mid).
Without continuum, the maximum deviation of the observed and modeled effective height is as large as $14\rm\,km$, and the mean residual is about twice as large.
Replacing the CKD water continuum data with more recent data of the MT-CKD continuum (version 2.5) has only a little effect on the spectra (not shown), i.e.\ the mean residuum is slightly changed from $0.462$ to $0.467\rm\,km$. 


\subsubsection{Impact of temperature and pressure}
\label{sssec:pT}

Transmission spectroscopy is primarily used for composition retrieval and is thought to have only little value for temperature retrieval \citep{Madhusudhan14,Heng17t}.
On the other hand, for the operational data processing of ACE-FTS observations temperature is first retrieved from the relative and absolute intensity of \chem{CO_2} lines \citep{Bernath17}.
However, the ACE-FTS data processing exploits about 30 spectra of a limb scan individually to infer the profile information on temperature (and gas abundances), whereas here we use a kind of mean transmission spectrum to resemble exoplanet remote sensing.
Nevertheless it is instructive to assess the temperature sensitivity of the effective height spectra.

The ACE-FTS arctic winter atlas comprises measurements of December to February in the 60 to $90^\circ$N latitude belt.
Accordingly we have recalculated the effective height (with all 38 molecules included) with the SAW temperature \citep{Anderson86} replaced by the fifteen arctic winter CIRA temperature profiles \citep{Fleming90}.
The resulting residua means span the 0.448 to $0.522\rm\,km$ interval with the smallest mean for the $60^\circ$N February CIRA profile (the corresponding norm is 73.47), i.e.\ with this profile the height is fitted slightly better than with the SAW profile.

Note that in addition to $33 \times 12$ temperature profiles CIRA also provides a single pressure profile as function of altitude.
This profile significantly differs from the SAW profile, it is about a factor of 1.46 larger around $40\rm\,km$ and a factor 0.65 smaller at ToA.
The effective height spectrum obtained with both CIRA pressure and temperature profiles shows larger deviations from the observed spectrum and is not considered further on.

As a further test of the temperature sensitivity of the effective height spectra the transmission has also been calculated with the other pressure and temperature profiles of the \citet{Anderson86} dataset.
For all five $p,~T$ profiles the residuals are significantly larger, the norms lie in the range $93$ to $178\rm\, km$.
If, in addition, the concentration profiles of \chem{H_2O}, \chem{O_3}, \chem{N_2O}, \chem{CH_4}, and \chem{CO} are also replaced, the deviations get even larger, with residua norms from $105$ to
$233\rm\,km$. Note that in all cases the maximum residuum is in the \chem{CO_2} band at $4.3\mue$. 
\qufig{fig:auxData} (bottom) shows effective heights and residuals for MLS with the largest mean difference of $1.715 \rm\,km$.
The results clearly confirm the choice of the SAW profile and emphasize the importance of choosing an appropriate pressure, temperature profile for fitting effective height observations.


\subsubsection{Assessment of the numerical quadrature}
\label{sssec:quad}

\begin{figure*}
 \centering\includegraphics[width=0.9\textwidth]{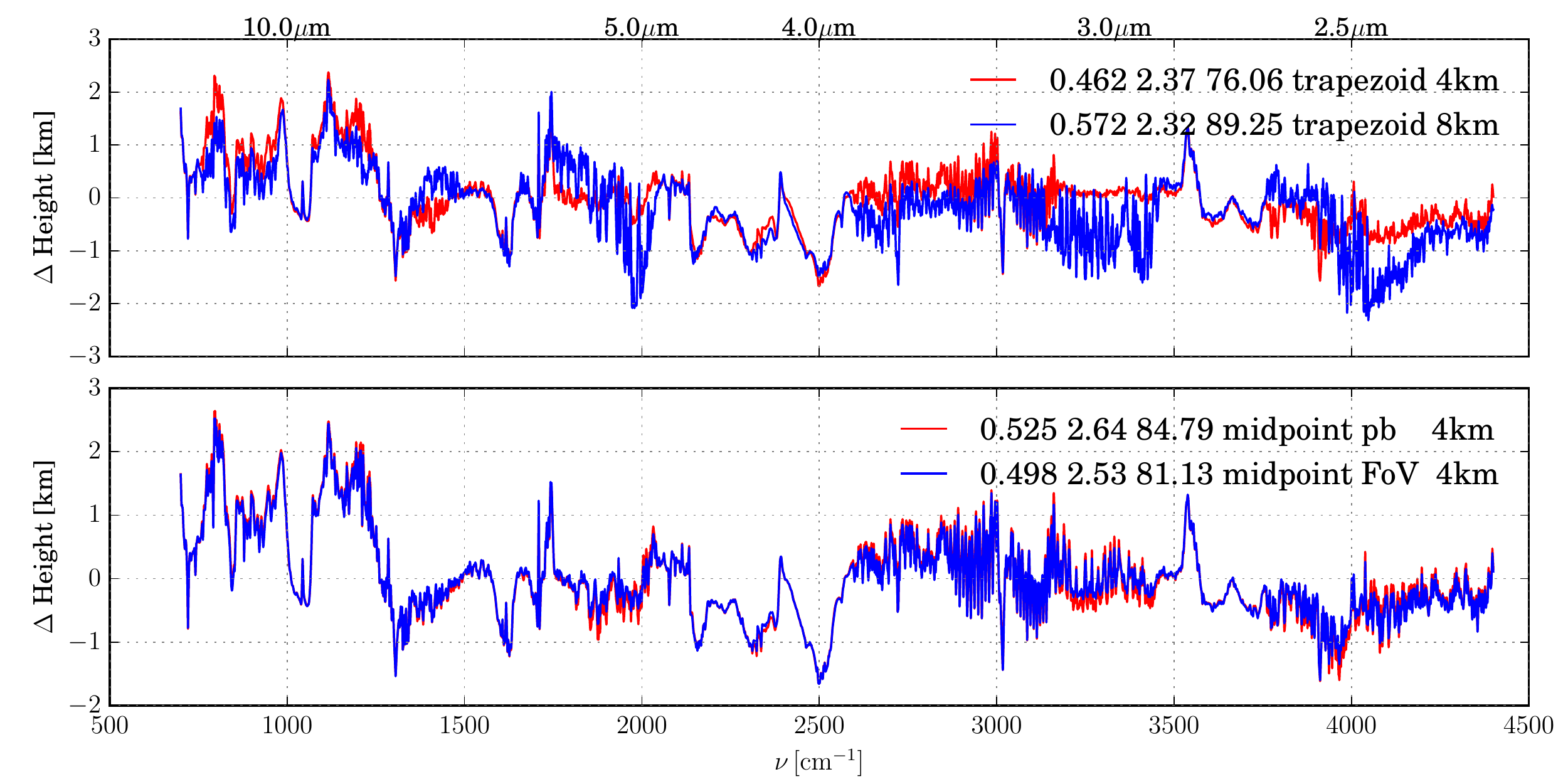}
 \caption{Comparison of quadrature methods for effective height residua spectra.
          Top: trapezoid ($\delta z_\text{t}=4\rm\,km$) vs.\ 8\,km.
          Bottom: midpoint quadrature with infinitesimal (pencil beam) vs.\ finite field-of-view ($\delta z_\text{t}=4\rm\,km$).
          ACE-FTS arctic winter; model spectra with 38 molecules as discussed in subsection 3.1.
          The numbers in the legend are the mean, maximum, and norm residuum of observation vs.\ model as defined in subsection \ref{ssec:detect}.}
 \label{fig:quadTrans}
\end{figure*}

The effective height spectra presented so far have been evaluated with the trapezoid quadrature \eqref{effHgtSum} given a series of GARLIC limb spectra with equidistant tangent points from 4 to $100\rm\,km$ in $\delta z_\text{t} = 4\rm\,km$ steps.
A denser tangent point grid does not significantly change the effective height: the difference of the $\delta z_\text{t}=4\rm\,km$ and $\delta z_\text{t}=2\rm\,km$ spectra is less than $0.4\rm\, km$, whereas a coarser
spacing of 8\,km gives substantial differences up to almost 2\,km, and the mean residuum \eqref{defMean} increases to $0.572\rm\,km$ (\qufig{fig:quadTrans} top).

The effective height integral \eqref{effHgt} approximated with the midpoint quadrature and a limb sequence with tangent points at $6, ~10, ~14, ~\dots,~98\rm\,km$ altitude also shows larger deviations to the observed
effective height, with a mean residuum of $0.525\rm\,km$.
Using a denser tangent grid spacing for the midpoint quadrature (i.e., $5, ~7, ~9, ~\dots,~99\rm\,km$) reduces this difference to $0.112\rm\,km$.

Note, however, that with midpoint quadrature contributions from the lowest atmospheric layers are missing if the individual transmission spectra are computed for pencil-beams.
Limb paths with different tangent heights probe different parts of the atmosphere, i.e.\ the temperature decreases by more than $10\rm\,km$ from $z_\text{t}=4\rm\,km$ to $z_\text{t}=6\rm\,km$,
and the ``effective'' path lengths through the atmosphere (ToA -- tangent -- ToA) is shortened by more than $20\rm\,km$ (for a pure geometric case ignoring refraction).
Modeling the limb transmissions with a rectangular field-of-view with half width $1\rm\,km$ improves the quality of the midpoint quadrature estimate (mean residuum $0.498\rm\,km$, see \qufig{fig:quadTrans} bottom).

\subsubsection{``Top-down'' analysis --- Impact of a missing molecule}
\label{sssec:topDown}

In subsection \ref{sssec:molecules} the relevance of a particular molecule has been inferred from its impact on the residual spectrum and estimated quantitatively by the decrease of the residual mean, maximum, or norm.  
An alternative approach is to study the increase of the residual due to the neglect of a single molecule in the list of absorbers.
Accordingly we have computed a series of effective height spectra with 37 molecules (denoted as $h_{37}$), i.e.\ with one of those gases excluded from the list of absorbers.

\qutab{tab:residuaGaussAWI2} reveals that the omission of \chem{H_2O}, \chem{CO_2}, \chem{O_3}, \chem{N_2O}, \chem{CH_4}, \chem{HNO_3}, or \chem{N_2} drastically increases the residuum norm.
Note that for \chem{HF} the residuum norm is slightly smaller, this might be a hint on inadequacies of the spectroscopic data and/or concentration profile.
The relative change of the residuum norm \eqref{relNormchange} can be as large as a factor ten, e.g.\ for carbon dioxide and ozone.
The importance of these molecules essentially confirms the results from the bottom-up analysis compiled in \qutab{tab:residuaGaussAWI1}.

The last column of \qutab{tab:residuaGaussAWI2} indicates that without \chem{H_2O}, \chem{CO_2}, \chem{O_3}, \chem{CH_4}, or \chem{HNO_3} the effective height spectrum changes by more than $10\rm\,km$.
Note that with all 38 molecules included in the model ($h_{38}$), the residuum spectrum $h_{38}-h_\text{obs}$ is ideally a noisy zero,
hence the maximum of $h_{38}-h_{37} \approx h_\text{obs}-h_{37} \equiv \Delta h_{37}$ is a measure of the strength of the neglected molecule's spectral feature.
This maximum residuum may overestimate the actual strength, but appears to be somewhat more objective than the commonly used difference of the center absorption and the neighboring ``continuum'' absorption.

\subsubsection{A preliminary list of important molecules}
\label{sssec:firstList}
\begin{figure*}
 \centering\includegraphics[width=0.9\textwidth]{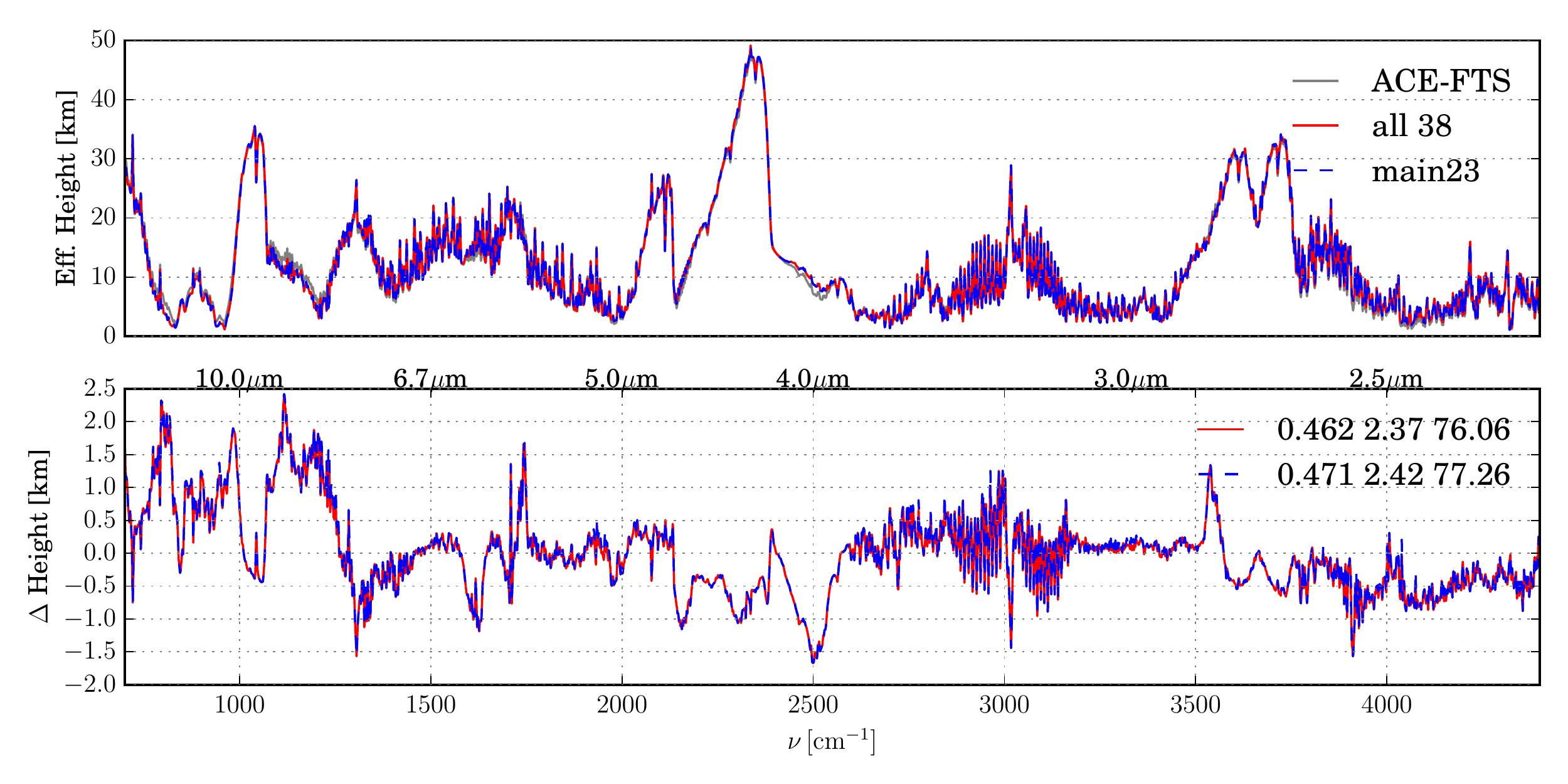}
 \caption{Comparison of transit spectra modeled with the 23 ``important'' gases or with all 38 gases. (Sub)-arctic winter, moderate resolution $\Gamma=1\cm$. Legend numbers as in \qufig{fig:residua_add}.}
 \label{fig:allORmain}
\end{figure*}

The bottom-up and top-down analysis summarized in the two tables can be used to define a (preliminary) list of important molecules for the Earth's transmission spectrum.
Clearly gases such as \chem{PH_3} or \chem{CClF_3} do not have any significant absorption and can be ignored.
Removing further species such as the heavy halogens also does not substantially increase the residuum.
\qufig{fig:allORmain} shows that with 23 absorbing molecules (i.e.\ ignoring 15 molecules) the mean and norm residuum increases only slightly to $0.471$ and $77.26\rm\,km$.
Note that also ignoring absorption of \chem{NH_3}, \chem{C_2H_6} or \chem{CHClF_2} would lead to an increase of the mean residuum by more than a hundred meters.

\subsection{The global view: weighted average effective height}
\label{ssec:global}

The study of the (sub)arctic winter atlas presented in the previous subsection has demonstrated the validity of our radiative transfer modeling approach, the impact of ``auxiliary'' data such as molecular
spectroscopy and atmospheric state parameters, and has allowed to define a list of relevant molecules.
Modeling effective height spectra for the (sub)arctic summer, the two midlatitude and the tropical case and comparing these with the corresponding ACE-FTS IR atlas leads to the same conclusions.
The mean or norm of the residuals as a function of the number of molecules modeled is essentially identical to \qutab{tab:residuaGaussAWI1} and confirms the list of ``important'' molecules given above.

When a planet is observed as a distant point source, it will likely be impossible to distinguish contributions from polar, midlatitude, or tropical regions.
Thus, when using the transit spectra generated from the ACE-FTS atlas as remote observations of an exoplanet, a disk-averaged spectrum can be approximated by combining one sixth of the arctic summer, arctic winter, midlatitude summer, midlatitude winter, and one third of the tropical spectrum.

\subsubsection{Global vs.\ seasonal/latitudinal atmospheric data}
For exoplanets little is known about seasonal and latitudinal climatologies, however, 3D model calculations have already been used successfully to study Earth-like exoplanets
\citep[e.g.][]{Shields14,Yang14,Godolt15,Turbet16,Kopparapu17,Wolf17}.
These advanced models can be used to construct a priori and or initial guess atmospheric state parameters for the retrieval from exoplanet observations.
In the following, however, we will only consider the six ``scenarios'' of the \citet{Anderson86} dataset (see subsection \ref{ssec:atmData}).
In particular, we will use the ``US Standard'' (USS) profile of the \citet{Anderson86} dataset for modeling the global effective height spectra in the following subsections.
As for nitric acid and the CFC's, the factor 3 enhancement of \chem{HNO_3} found for the arctic winter is not appropriate elsewhere, but increased CFC concentrations can be
observed in all cases.
Thus, the \chem{HNO_3}, \chem{CCl_3F} and \chem{CCl_2F_2} concentrations have been doubled henceforth.

\begin{figure*}[t]
 \centering\includegraphics[width=0.9\textwidth]{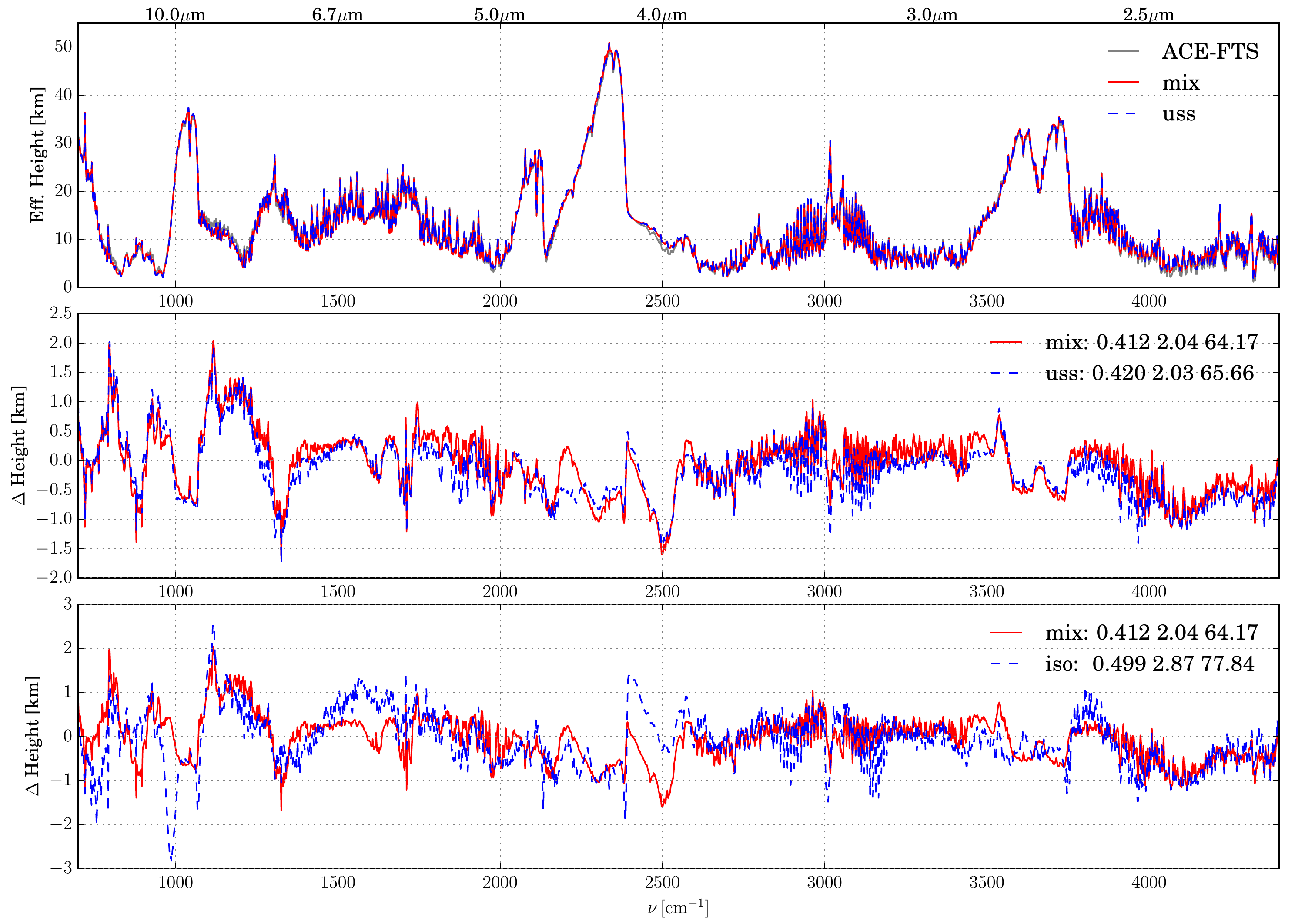}
 \caption{Comparison of global moderate resolution ($\Gamma=1\cm$) spectra: The ACE-FTS spectrum is a combination of the tropical, midlatitude, and arctic spectra.
          Model spectra (23 molecules) calculated as a mix with tropical, MLS, MLW, SAS, and SAW climatologies (red) or with the USS atmosphere (blue, top and center) or with an isothermal ($T=250\rm\,K$) atmosphere (blue, bottom, residuals only).  Legend numbers as in \qufig{fig:residua_add}.}
 \label{fig:globe_modRes_all_uss}
\end{figure*}

Before studying the detectability of various molecules, we compare the impact of atmospheric data on the disk-averaged spectrum.
As a reference we computed the tropical, midlatitude summer/winter and subarctic summer/winter spectra using the respective data (pressure, temperature, and the main IR absorbers) from \citet{Anderson86} and combined these with the weighting factors as described above similar to the combination of the ACE-FTS atlases.
In \qufig{fig:globe_modRes_all_uss} this spectrum is compared to a model spectrum obtained with the USS pressure, temperature, and concentration profiles indicating a slightly larger residuum mean and norm.
Moreover, we tested a constant temperature profile throughout the atmosphere (\qufig{fig:globe_modRes_all_uss} bottom) with $T=250\rm\,K$.
The discrepancies between observed and model spectrum are clearly visible, and the mean and norm residuals are increased by roughly 20\%. 
For a discussion of transmission spectra of hot Jupiter isothermal atmospheres see \citet{Heng17t}.

In view of the little improvement and the higher computational cost of the ``mix'' scenario compared to the USS scenario, we will henceforth use the USS profiles to compute the effective height spectra.

\subsubsection{Moderate resolution spectra}
\label{sssec:modRes}

\begin{table*}[t]
 \caption{\label{tab:residuaGlobal} %
  Norm \eqref{defNorm}, relative change of norm \eqref{relNormchange}, maximum change $\max(\delta h) = \max(h_{23}-h_{22})$ of effective height \eqref{effHgt}, and the change of the additional transit depth \eqref{transDepth} due to omission of a single molecule.
The norm for the spectra with all 23 molecules included is given in the very first row.
 Note that the $\delta h$ columns give the maximum change, e.g.\ the peak at $1040\cm$ for ozone and at $3030\cm$ for methane, compare \qufig{fig:missingModRes} and \qufig{fig:missingLowRes}.}
 \begin{tabular}{l|rrrr|rrrr}
 \hline
         & \multicolumn{4}{c}{moderate resolution} & \multicolumn{4}{c}{low resolution} \\
         &    norm    & relNorm & $\delta h$ & $\Delta \delta d_\text{t}$ &    norm    & relNorm & $\delta h$ & $\Delta \delta d_\text{t}$ \\
         &    [km]    &         & [km]       &            &    [km]    &         & [km]       &    \\
 \hline
         &     65.656 &          &          &            &     19.279 &          &          &            \\
 CO2     &    815.414 &   11.419 &   36.386 &    9.6e-07 &    253.121 &   12.129 &   34.364 &    9.1e-07 \\
 O3      &    618.729 &    8.424 &   34.966 &    9.2e-07 &    189.244 &    8.816 &   32.519 &    8.6e-07 \\
 H2O     &    509.956 &    6.767 &   19.296 &    5.1e-07 &    152.659 &    6.918 &   12.541 &    3.3e-07 \\
 CH4     &    330.335 &    4.031 &   22.055 &    5.8e-07 &     93.601 &    3.855 &   10.393 &    2.7e-07 \\
 N2O     &    159.233 &    1.425 &    8.950 &    2.4e-07 &     47.058 &    1.441 &    8.069 &    2.1e-07 \\
 N2      &    139.539 &    1.125 &    8.046 &    2.1e-07 &     42.818 &    1.221 &    6.616 &    1.7e-07 \\
 HNO3    &    114.697 &    0.747 &    8.110 &    2.1e-07 &     33.768 &    0.752 &    5.636 &    1.5e-07 \\
 O2      &     86.748 &    0.321 &    4.130 &    1.1e-07 &     25.622 &    0.329 &    3.240 &    8.5e-08 \\
 CCl2F2  &     83.632 &    0.274 &    4.615 &    1.2e-07 &     24.261 &    0.258 &    2.919 &    7.7e-08 \\
 CCl3F   &     72.621 &    0.106 &    4.535 &    1.2e-07 &     20.804 &    0.079 &    2.455 &    6.5e-08 \\
 NO2     &     70.396 &    0.072 &    2.427 &    6.4e-08 &     20.760 &    0.077 &    1.850 &    4.9e-08 \\
 ClONO2  &     66.968 &    0.020 &    0.873 &    2.3e-08 &     19.649 &    0.019 &    0.505 &    1.3e-08 \\
 N2O5    &     66.835 &    0.018 &    1.000 &    2.6e-08 &     19.637 &    0.019 &    0.731 &    1.9e-08 \\
 CHClF2  &     66.532 &    0.013 &    0.198 &    5.2e-09 &     19.564 &    0.015 &    0.116 &    3.1e-09 \\
 CO      &     64.777 &    0.013 &    2.383 &    6.3e-08 &     18.692 &    0.030 &    1.098 &    2.9e-08 \\
 C2H6    &     65.913 &    0.004 &    0.307 &    8.1e-09 &     19.344 &    0.003 &    0.132 &    3.5e-09 \\
 CF4     &     65.831 &    0.003 &    1.042 &    2.7e-08 &     19.261 &    0.001 &    0.350 &    9.2e-09 \\
 OCS     &     66.122 &    0.007 &    1.090 &    2.9e-08 &     19.361 &    0.004 &    0.720 &    1.9e-08 \\
 CH3Cl   &     65.718 &    0.001 &    0.109 &    2.9e-09 &     19.287 &    0.000 &    0.057 &    1.5e-09 \\
 HOCl    &     65.691 &    0.001 &    0.028 &    7.3e-10 &     19.290 &    0.001 &    0.019 &      5e-10 \\
 NH3     &     65.711 &    0.001 &    0.119 &    3.1e-09 &     19.297 &    0.001 &    0.026 &      7e-10 \\
 NO      &     65.630 &    0.000 &    0.311 &    8.2e-09 &     19.269 &    0.001 &    0.193 &    5.1e-09 \\
 SO2     &     65.655 &    0.000 &    0.065 &    1.7e-09 &     19.278 &    0.000 &    0.035 &    9.2e-10 \\
 \hline
 \end{tabular}
\end{table*}

\begin{figure*}
 \centering\includegraphics[width=0.9\textwidth]{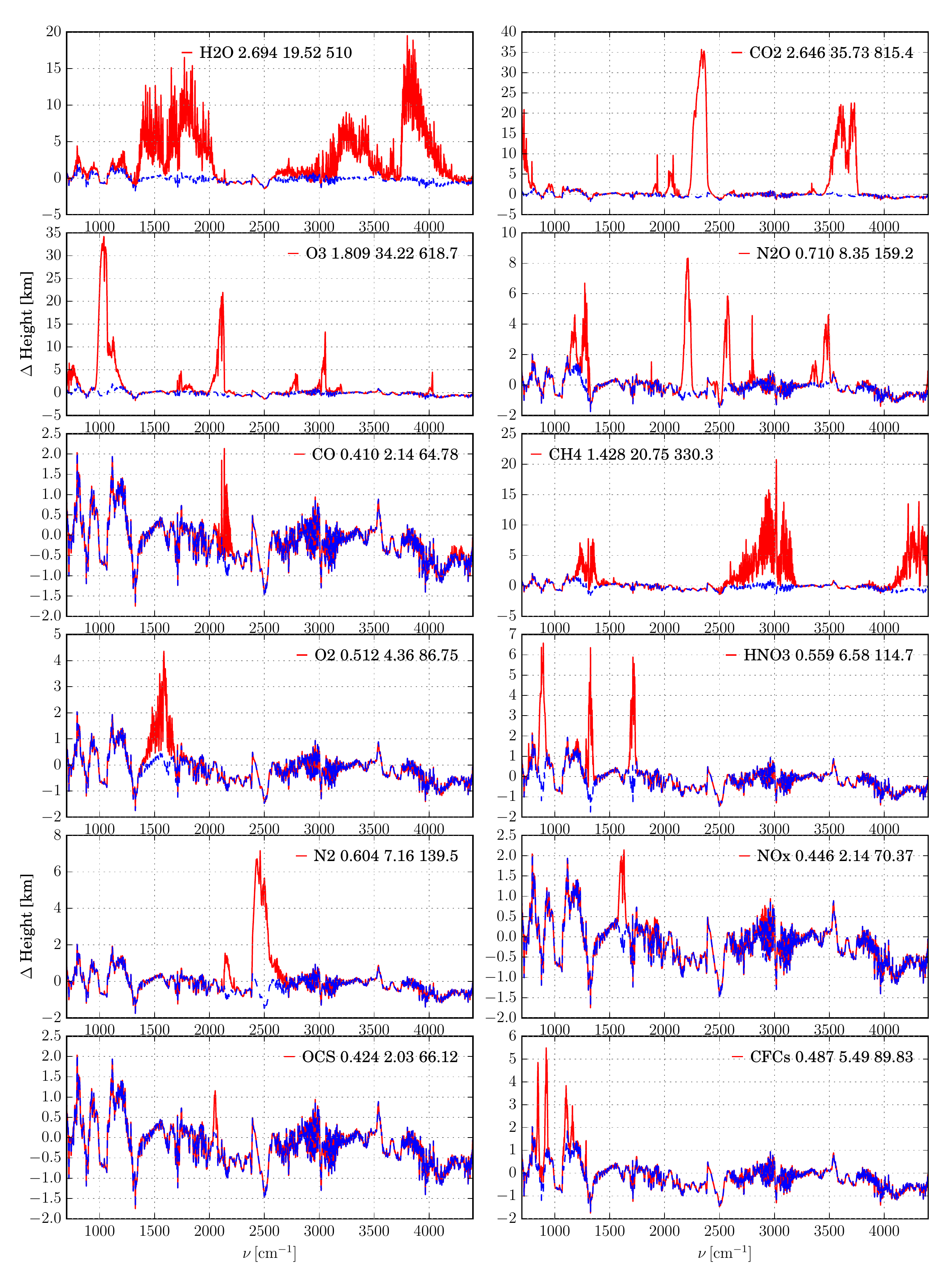}
 \caption{Impact of a missing species on global residual spectra (moderate resolution $\Gamma=1\cm$).
          The bottom right plot shows \chem{CCl_3F}, \chem{CCl_2F_2}, and \chem{CF_4} combined.
          For the reference spectrum (blue) with 23 molecules the residuum mean and norm are $0.42\rm\,km$ and $65.6\rm\,km$, respectively (see \qutab{tab:residuaGlobal}).
          Legend numbers as in \qufig{fig:residua_add}.}
 \label{fig:missingModRes}
\end{figure*}

The impact of a missing molecule on the model spectrum is listed in \qutab{tab:residuaGlobal}, and residual spectra are shown in \qufig{fig:missingModRes}.
Molecules that were prominent in the (sub)arctic spectra are also prominent for the global view.
In particular, the relative change of the residuum norm due to the neglect of carbon dioxide is greater than ten, and \chem{H_2O}, \chem{O_3}, \chem{N_2O}, \chem{CH_4}, \chem{N_2}, and \chem{N_2O} increase the norm by more than a factor two.
Note that the ranking of the relative norm changes in \qutab{tab:residuaGaussAWI2} and \ref{tab:residuaGlobal} is identical for the strongest absorbers, indicating a similar molecular impact for the global and arctic case.

These large changes of the residuum norm are also clearly visible as differences between the effective height spectra, that become larger than $30\rm\,km$ for \chem{CO_2} and \chem{O_3} (third numeric column in \qutab{tab:residuaGlobal}).
Note that for some gases (\chem{CO}, \chem{NO}, \chem{CF_4}) the exclusion leads to slightly smaller residual norms, what might be attributed to imperfect concentration profiles or spectroscopic data.
Furthermore note that the $\max(\delta h)$ given in \qutab{tab:residuaGlobal} are similar to the maximum residual $\max(\Delta h)$ given in the legend of \qufig{fig:missingModRes} (see the
discussion in subsection \ref{sssec:topDown}).

Molecules that do not show a significant impact on the residua are candidates for a further consolidation of the list of important absorbers:
Neglecting \chem{C_2H_6}, \chem{CF_4}, \chem{OCS}, \chem{CH_3Cl}, \chem{HOCl}, \chem{NH_3}, \chem{NO}, or \chem{SO_2} leads to a relative increase of the residuum norm of less than one percent.
Note, however, that \chem{OCS}, \chem{NO}, and \chem{CF_4} change the effective height spectrum by about $1.1$, $0.3$, and $1.0\rm\,km$, respectively (see \qufig{fig:missingModRes} bottom).

\subsubsection{Low resolution spectra}
\label{sssec:lowRes}

The observed effective height analyzed so far has been obtained by convolution of the high resolution ACE-FTS spectra with a Gauss response function of half width $\Gamma=1.0\cm$, and the model spectra have been degraded analogously. 
For an assessment of the relevance of molecular species with lower resolution, both the ACE-FTS and GARLIC spectra have been convolved with a Gauss of $\Gamma=10\cm$.

\begin{figure*}
 \centering\includegraphics[width=0.9\textwidth]{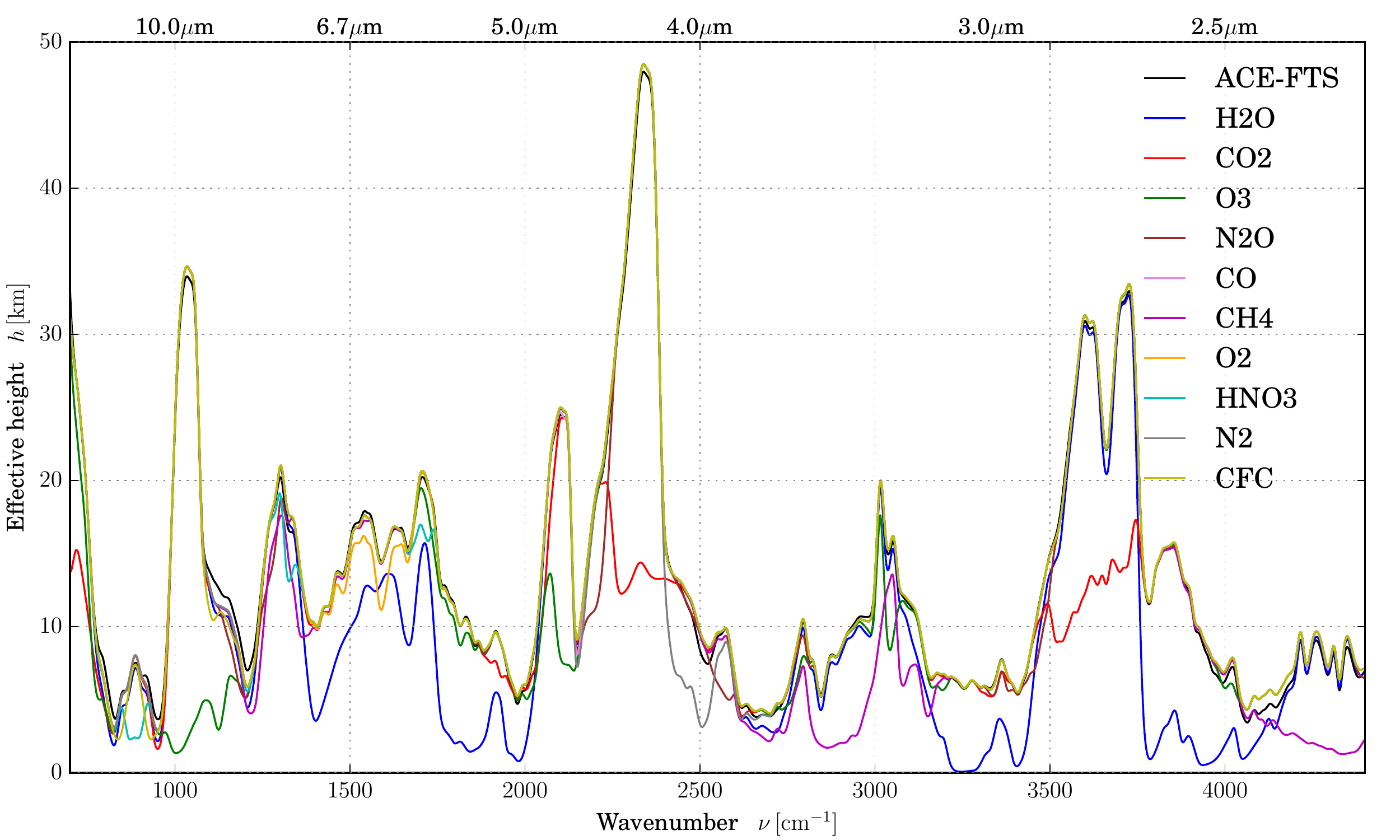}
 \caption{Impact of a missing species on the global effective height spectrum (low resolution $\Gamma=10\cm$).}
 \label{fig:missingLowResEH}
\end{figure*}

The effective height spectra for all but one molecule are depicted in \qufig{fig:missingLowResEH}.
As already discussed in the previous subsection \ref{sssec:modRes}, leaving out molecules such as carbon dioxide or ozone does not reduce the effective height to zero.
Because of other interfering gases in the $4.3 \rm\,\mu m$ band, excluding carbon dioxide reduces the effective height from $48 \rm\,km$ to about $10 \text{--} 20\rm\,km$.
Likewise, in the $9.6\rm\,\mu m$ region the absence of ozone in the model reduces $h$ to roughly $5 \rm\,km$.
The strong absorption features of these gases are superimposed on a non-negligible background contribution of interfering gases, e.g.\ water vapor, which will make the quantification of the
molecules abundance difficult.

\begin{figure*}
 \centering\includegraphics[width=0.9\textwidth]{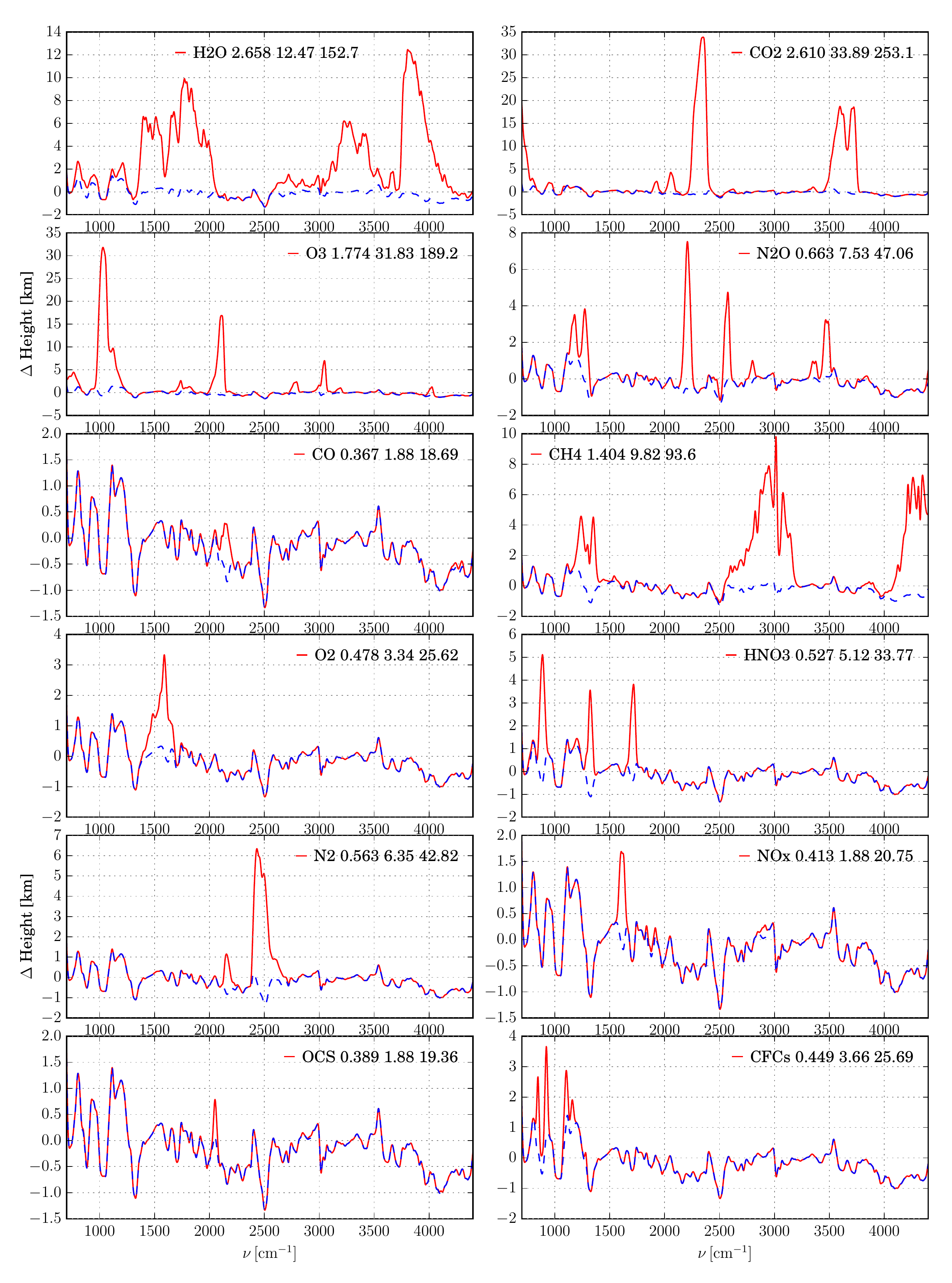}
 \caption{Effective height residual spectra due to the exclusion of a single molecule (low resolution $\Gamma=10\cm$, see also \qufig{fig:missingModRes}).
          For the reference spectrum (blue) with 23 molecules the residuum mean and norm are $0.38\rm\,km$ and $19.3\rm\,km$, respectively (see \qutab{tab:residuaGlobal}).
          Legend numbers as in \qufig{fig:residua_add}.}
 \label{fig:missingLowRes}
\end{figure*}

The last four columns of \qutab{tab:residuaGlobal} quantify the impact of an ignored molecule on effective height residuals and transit depths.
Note that the residual norms are much smaller for the low resolution case (roughly a factor $\sqrt{10} \approx 3$ corresponding to the length of the data vector), but the relative change of the norm is similar and the ranking of the eleven largest changes is identical for both resolutions.
The main IR absorbers \chem{H_2O}, \chem{CO_2}, \chem{O_3}, and \chem{CH_4} have a drastic effect on the spectra (\qufig{fig:missingLowRes}) and their omission significantly increases the residuum norm.
Excluding nitrous oxide, oxygen, nitric acid, and nitrogen also lead to marked increases of the residual.
As for the moderate resolution case (\qufig{fig:missingModRes}), removing carbon monoxide reduces the residuum mean and norm only slightly, but can be clearly seen in the spectrum.

The importance of including heavy species, esp.\ some CFCs, in the modeling is also evident here.
Along with \chem{HNO_3}, both \chem{CCl_3F} and \chem{CCl_2F_2} significantly contribute to the absorption around $900\cm$, see \qufig{fig:missingLowRes} (bottom right).
In this spectral region the Gaussian width of $\Gamma=10\cm$ is equivalent to a resolution $\nu / \Gamma \approx 100$,
i.e.\ the visible impact of these CFCs on the effective height confirms the statement of \citet{Schneider10} on the detectability of these ``technosignatures''.

Despite the significantly worse (factor 10) resolution, the maximum change of the effective height $\delta h$ is only slightly smaller for \chem{CO_2} and \chem{O_3}, hence the transit depth change is still close to $1\rm\,ppm$.
Smearing of the fine peak structure of methane reduces $\delta h$ by more than a factor two, whereas for \chem{H_2O}, \chem{N_2O}, \chem{O_2}, \chem{N_2}, \chem{HNO_3}, and the CFC's the reduction is less than a factor two.

\subsubsection{Signal-to-Noise Ratios}
\label{sssec:snr}

For an estimate of the SNR and its dependence on resolution further runs with response function half widths $\Gamma=2, ~5, ~20$ and $50\cm$ have been performed.
Note that for \chem{HNO_3} the absorption peak around $890\cm$ does not show up anymore for $\Gamma=50\cm$.
Given the effective height spectrum of all 23 molecules and the spectrum of 22 molecules the change of the additional transit depth is estimated by
$ \max\bigl(d_\text{t,23} - d_\text{t,22} \bigr)$ (see \qutab{tab:residuaGlobal}).
For the $\rm SNR_T$ according to \qeq{snr} we assume (similar to \citet{Hedelt13}, see his Table 3) a James Webb Space Telescope (JWST) configuration with $A=240\rm\,m^2$ and $q=0.15$,
a stellar radius and temperature as for the Sun ($6.96\cdot 10^5\rm\,km$ and $5770\rm\,K$), an integration time of $12.98\rm\, h$ for the Sun-Earth system, and an observer-star distance of $10\rm\,pc$.
The resolving power is estimated by the FWHM (full width at half maximum) of the Gaussian spectral response function at the band center, i.e.\ $R = \nu_\text{band}/2\Gamma$.

\begin{figure*}[t]
 \centering\includegraphics[width=0.9\textwidth]{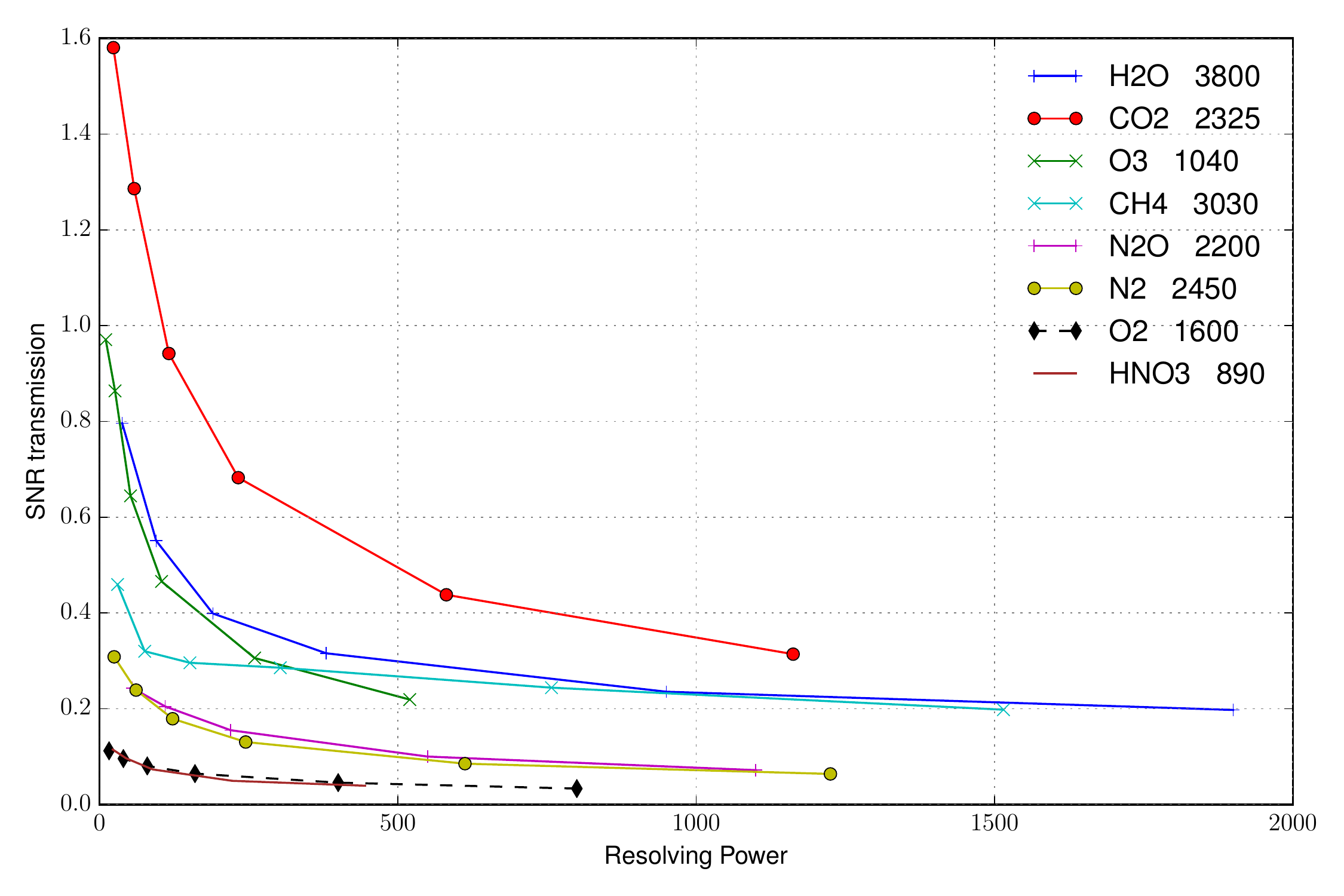}
 \caption{Single-to-noise ratios of Earth's transmission spectrum seen from a 10\,pc distance. The numbers in the legend indicate the band center wavenumber in $\cm$.}
 \label{fig:snr}
\end{figure*}

\qufig{fig:snr} depicts the SNRs for the strongest molecules.
An SNR better than one can only be expected for carbon dioxide with a resolving power of 100 or smaller.
Because the SNR is proportional to the inverse distance and the square root of time, $\text{SNR} \sim \sqrt{t}/D$, reasonable SNRs may also be possible for ozone when an exo-Earth at a few parsecs is observed for multiple transits with $R \le 20$.
Note, however, that ``simply'' co-adding spectra recorded at multiple transits might be challenging because of the temporal variability of the host star
or the degradation of the detector performance.
In this respect, a planet orbiting closer to its host star allowing more observations in a shorter period of time would be advantageous.
For example, an Earth-like planet at 1\,AU could only be observed five times during JWST's lifetime, leading to a factor $\approx\! 2$ improvement of the SNR.
Furthermore, detection of water vapor or methane might be feasible under favorable circumstances.

\subsubsection{Clouds}
\label{sssec:clouds}

Clouds cover more than half of Earth's surface and play a dominant role in the climate system \citep{Stubenrauch13}.
Although the ACE-FTS atlases have been compiled by averaging cloud-free spectra only, it is nevertheless instructive to exploit these data for an assessment of the impact of clouds.
Following \citet{GarciaMunoz12e} clouds are assumed to block the radiation traversing the lower most limb ray at $4\rm\, km$, i.e.\ the very first transmission contributing to the integral/sum in \qeqs{effHgt} and \eqref{effHgtSum} is set to  $\T(\nu,z_1)=0$.
(Note that ``typical'' water clouds can be expected for somewhat lower regions, whereas ice clouds are typically found in altitudes around/up to the tropopause.)
Furthermore we assume a 50\% cloud cover for each of the 5 latitude bands / seasons, so the final spectra are given by the mean of the cloud-free and fully cloud-contaminated global spectra.

\begin{figure*}
 \centering\includegraphics[width=0.9\textwidth]{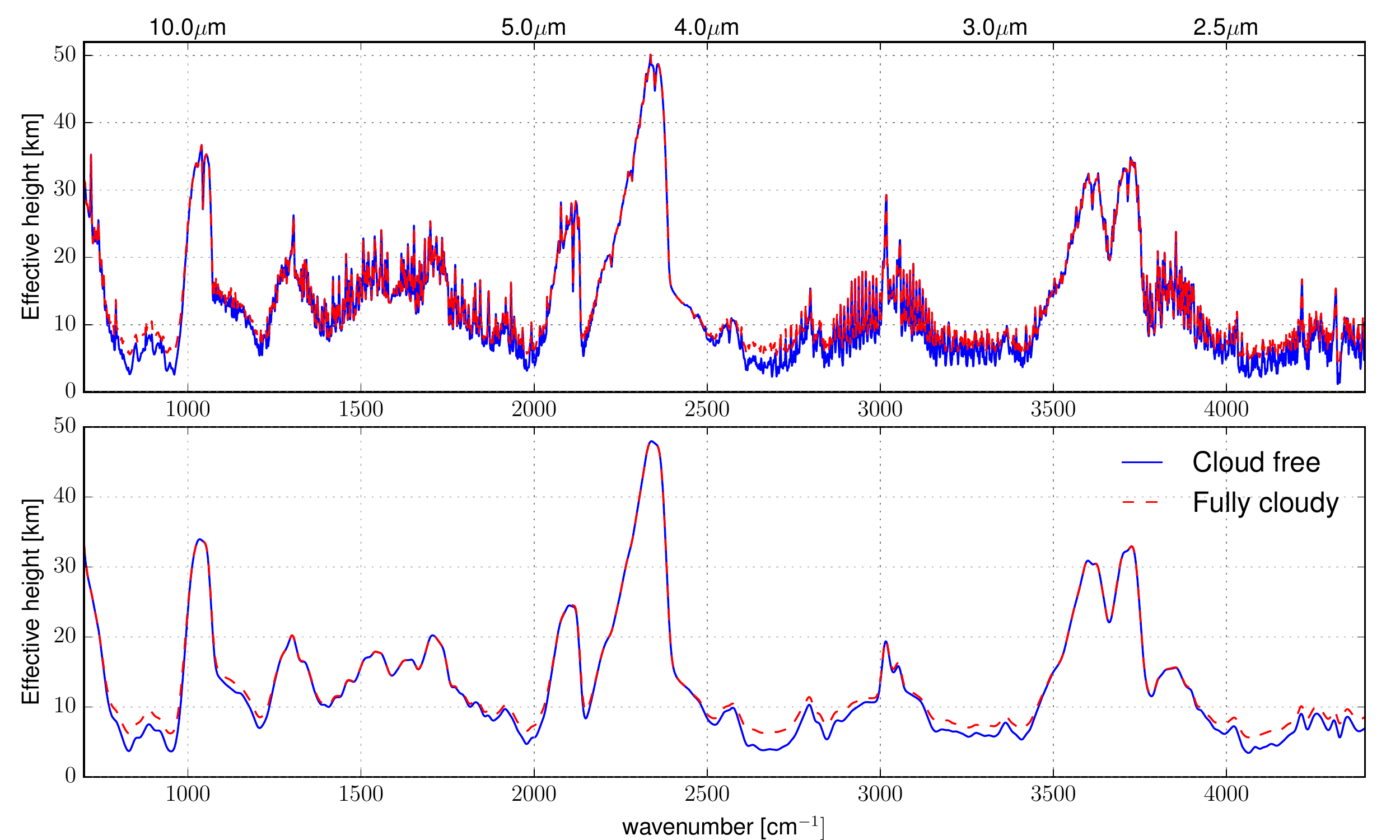}
 \caption{Impact of clouds on global effective height spectrum (top: moderate resolution $\Gamma=1\cm$, bottom: low resolution $\Gamma=10\cm$)}
 \label{fig:clouds}
\end{figure*}

\qufig{fig:clouds} demonstrates that a cloud layer changes the global effective height spectrum essentially for small heights, whereas spectral regions characterized by strong absorption due to, e.g., carbon dioxide or ozone are not affected.
Accordingly one can expect that clouds do not significantly change the detectability of these strong absorbers.


\section{Discussion}
\label{sec:discussion}

Our analysis has demonstrated that ten gases substantially contribute to Earth's transit spectrum as observed by the ACE-FTS instrument:
\chem{CO_2}, \chem{O_3}, \chem{H_2O}, \chem{CH_4}, \chem{N_2O}, \chem{N_2}, \chem{HNO_3}, \chem{O_2}, and the CFCs \chem{CCl_3F} and \chem{CCl_2F_2} ordered according to the relative change of the residual norm (see \qutab{tab:residuaGlobal}).
However, the transit spectrum modeled with these 10 ``main'' gases only results in large residua esp.\ around $1630\cm$, that can be attributed to \chem{NO_2}.
The neglected absorption due to \chem{NO} is especially seen in the low resolution spectrum around $1900\cm$.
Carbon monoxide has a substantial impact at $2150\cm$ (and much smaller around $4300\cm$).
Furthermore, the omission of \chem{OCS} leads to noticeable residuals at $2070 \cm$, and \chem{CF_4} is visible around $1285\cm$.
\chem{ClONO_2} clearly shows up at $800$ and $1745\cm$ (the band around $1300\cm$ is less prominent in the spectra), and \chem{N_2O_5} at $750$ and $1240\cm$.
Hence, adding nitrogen oxides, carbon monoxide, carbonyl sulfide, tetrafluoromethane, chlorine nitrate, and nitrogen pentoxide in the model removes these discrepancies, although these five species do not alter the mean and norm residual.
In conclusion, the preliminary list of ``important'' absorbers defined in subsection \ref{sssec:topDown} can therefore be further shortened by deletion of \chem{C_2H_6}, \chem{CH_3Cl},
\chem{CHClF_2}, \chem{HOCl}, \chem{NH_3}, and \chem{SO_2}.
 
\begin{figure*}[t]
 \centering\includegraphics[width=0.9\textwidth]{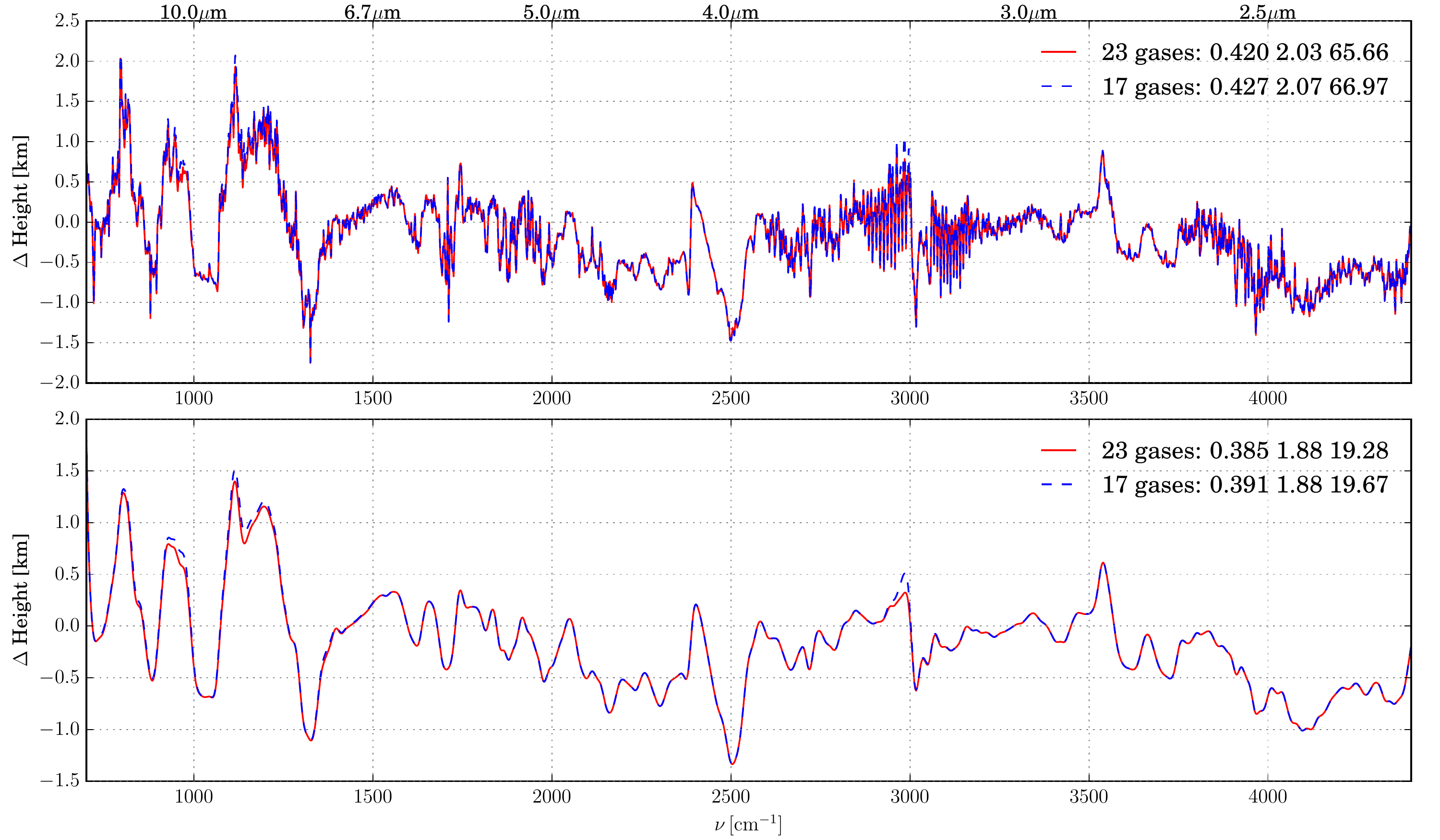}
 \caption{Comparison of transit spectra residuals modeled with the 23 gases (see \qutab{tab:residuaGlobal}) or with 17 gases:
          \chem{H_2O}, \chem{CO_2}, \chem{O_3}, \chem{N_2O}, \chem{CO}, \chem{CH_4}, \chem{O_2}, \chem{NO}, \chem{NO_2}, \chem{HNO_3}, \chem{OCS}, \chem{N_2}, \chem{CCl_3F}, \chem{Cl_2F_2}, \chem{CF_4}, \chem{ClONO_2}, \chem{N_2O_5}. 
          Moderate resolution $\Gamma=1\cm$ (top) and low resolution $\Gamma=10\cm$ (bottom).
          Legend numbers as in \qufig{fig:residua_add}.}
 \label{fig:allMain}
\end{figure*}

The comparison of spectra modeled with the 17 ``main'' gases and with 23 molecules is shown in \qufig{fig:allMain}.
Although the residual is somewhat larger, the selection of the relevant molecules is especially important for the quantitative estimate of concentrations using inversion techniques:
First, ignoring irrelevant species helps to speed up the forward modeling;
secondly, limiting the number of unknown gas concentrations, i.e.\ reducing the size of the state vector, will lead to a better conditioning of the inverse problem.
Nevertheless, neglecting further gases in the modeling leads to a visible increase of the low or moderate resolution residual spectra.
However, the importance of these gases for the modeling does not necessarily imply their detectability in noisy low resolution spectra.

Except for \chem{CO}, \chem{NO}, \chem{OCS}, \chem{CF_4}, \chem{ClONO_2}, and \chem{N_2O_5} our list is identical to the eleven ``spectroscopically most significant molecules'' used by \citet{Kaltenegger09}.
\citet{Rugheimer13} modeled Earth's reflection and emission spectra using ``21 of the most spectroscopically significant molecules''.
For the validation of the SMART code \citep{Meadows96} with AIRS \citep{Chahine06} and ATMOS \citep{Abrams96a} observations eight species (\chem{H_2O}, \chem{CO_2}, \chem{O_3}, \chem{N_2O},
\chem{CO}, \chem{CH_4}, and \chem{O_2}) have been included in the modeling by \citet{Robinson11} and \citet{Misra14}, respectively.
\citet{Barstow16} considered \chem{H_2O}, \chem{CO_2}, \chem{O_3}, \chem{CO}, \chem{CH_4}, \chem{O_2}, \chem{SO_2}, \chem{OCS}, \chem{N_2} for retrieval tests of exo-Earths and exo-Venuses, but
found \chem{SO_2} and \chem{OCS} to be negligible for Earth.
Also note that our list of seventeen gases is appropriate for Earth and may change for other planets.

The limited spectral resolution can mask the molecular absorption features, and we have assessed this by convolution of the high resolution ACE-FTS spectra and monochromatic model spectra with a Gaussian response function.
In particular, we have assumed that the width of the Gaussian is constant over the entire spectral domain.
Note that for modeling the ACE-FTS spectra the monochromatic transmission \eqref{beer} has to be convolved with an instrument line shape given approximately by $2L \, \sinc(2\pi L \nu)$ where $L$ is the maximum optical path difference ($L=\pm 25\rm\,cm$ for ACE-FTS).

In contrast to the constant width FTS, instruments for exoplanet observations are frequently characterized by a constant resolving power $R = \nu / \delta\nu$.
The InfraRed Spectrograph (IRS) of the Spitzer Space Telescope had two modules for moderate and low resolution in the $5 \TO 38\mue$ region where the moderate resolution $R = 600$ corresponds to our HWHM $\Gamma=1\cm$ at the low wavenumber end.
The IR and NIR spectrographs of the ARIEL (Atmospheric Remote sensing Infrared Exoplanet Large survey) ESA mission candidate aim to observe the $1.95 \TO 7.8\mue$ interval with a resolving
power of $R=100$ and the $1.25 \TO 1.95 \mue$ interval with $R=10$, respectively \citep{Tinetti16}.
The Mid InfraRed Instrument (MIRI) on the James Webb Space Telescope (JWST) will offer two spectrometer modes, the low resolution spectrograph with $R=100$ in $5 \,\text{--}\,12\mue$ and the medium resolution spectrometer with $R=1300 \TO
3700$ in $5 \TO 28\mue$, and JWST's Near-InfraRed Spectrograph (NIRSpec) covers four wavelength regions up to $5.2\mue$ with medium ($R=1000$) or high ($2700$) resolution \citep[e.g.][]{Stevenson16}.

Our spectra (both observed and modeled) are in agreement with other transit spectra published before.
The maximum effective height in the \chem{CO_2 ~ \nu_3} band around $2340\cm$ (about $48\rm\,km$ for the moderate resolution and $\approx 46\rm\,km$ for the low resolution) is compatible with spectra modeled by \citet{Misra14} and \citet{Kaltenegger09}.
The effective height in the center of the ozone band of about $35\rm\,km$ is somewhat lower than shown in \citep{Kaltenegger09},
and the minimum effective height of roughly $1 \TO 4 \rm\,km$ (depending on resolution) is considerably smaller here, both for the observed and model spectrum (presumably because of the absence of clouds).
The ``atmospheric radius'' of an Earth-like planet shown by \citet{Meadows17} has a maximum of about $60\rm\,km$ in the $\nu_2$ and $\nu_3$ bands of \chem{CO_2}, a similar peak at $9.6\mue$ at the ozone band, and the minimum radius is larger than observed by ACE-FTS.
The maximum of the effective tangent height around the \chem{CO_2} $4.3\mue$ band shown by \citet{Rauer11} is somewhat lower, whereas the maximum around the \chem{O_3} $9.6\mue$ band is larger than ours.

The SNR's reported here are also comparable with those given by \citet{Rauer11} (Table 3).
Differences can be attributed to, e.g., the atmospheric setup (here ``US Standard'' \citep{Anderson86} with enhanced carbon dioxide, methane, and nitric acid concentrations and additional trace gases), spectroscopic database (HITRAN\,2016 vs.\ 2004, see also remark below), and the actual estimate of the effective height changes.
Furthermore note that the analysis of subsection \ref{sssec:snr} only provides a rough SNR estimate assuming an ideal detector and non-variable sources;
for an extended noise model see, e.g., \citet{Hedelt13}.

In our analysis we have used atmospheric profiles as given by standard Earth climatologies \citep{Anderson86,Fleming90,Remedios07}.
Only in a few cases, where the default was either clearly inadequate (outdated \chem{CO_2} and \chem{CH_4} concentrations) or where discrepancies between observation and model were obvious (\chem{HNO_3} and CFC 11 and 12), profiles were adjusted by visual inspection of the spectra.
Fitting atmospheric state parameters (e.g.\ concentration scaling factors) by means of numeric optimization techniques is clearly advantageous and this study serves as a preparation for an analysis of the ACE-FTS spectra with nonlinear least squares inversion \citep[cf.][]{GimenoGarcia11}.
Although transmission spectra are primarily used for concentration retrievals, the analysis of subsection \ref{sssec:pT} has indicated the
importance of pressure and temperature profiles that are hardly available for exo-Earths; clearly the retrieval of $p$ and $T$ will make the inverse
problem even more challenging.  Note that for the data analysis of the ACE-FTS spectra pressure, temperature, and the \chem{CO_2} mixing ratio are
determined in a first step before the trace gas concentration retrieval \citep{Bernath17}.  

Success or failure of the inversion critically depends on the proper initial guess or a priori atmospheric profiles.
The temperature profile has been shown to impact the effective height (cf.\ subsection \ref{sssec:pT}), on the other hand \qufig{fig:effHeightObs} shows a surprisingly little spread of the observed spectra for the five seasons/latitudes.
Nevertheless, the choice of appropriate temperatures is clearly an important issue for the analysis of the ACE-FTS spectra or, more generally, planetary transmission spectra.
Among all IR relevant atmospheric species water has the highest variability (esp.\ in the troposphere), however, according to \qufig{fig:effHeightObs} this is only partly propagated into the effective height spectra near its band centers and the choice of the \chem{H_2O} profile might be less critical for the performance of the fitting.

Due to the mature quality of the current spectroscopic databases, the modeled effective height is essentially independent of the choice of HITRAN or GEISA.
However, using the much older HITRAN\,86 database with 233 thousand lines of 28 molecules significantly increases the mean and norm residual to $0.81\rm\,km$ and $154\rm\,km$ (compare the $m=28$ row in
\qutab{tab:residuaGaussAWI1}), with large peaks up to almost 10\,km in the residuum spectrum in the center of the methane bands.
Moreover, the version of the continuum did not change the model spectrum notably (the \chem{H_2O} continuum is especially important in the lower troposphere).
\citet{Esposito07} noted that ``typical differences introduced by the two \chem{H_2O} continuum models are of one order of magnitude less than typical differences arising from different line parameters.''
In conclusion, the choice of spectroscopic input data is not expected to impact the analysis substantially.
However, this may be different for water-rich and hot atmospheres \citep[e.g.][]{Bailey09,Yang16}.

Our main objective has been to identify molecular signatures in Earth's transit spectra useful for atmospheric retrieval of terrestrial exoplanets.
Alternatively, one can view this study as a validation of the GARLIC forward model.
A similar validation study for the SMART code \citep{Meadows96} using ATMOS observations \citep{Abrams96a} has been presented by \citet{Misra14}.
However, the approach used here appears to be suboptimal for a thorough validation:
First, it would be more appropriate to compare model and observed transmission spectra corresponding to individual tangent heights (similar to \citet{Kaltenegger09,Misra14}).
Secondly, the comparison should be performed for high spectral resolution of the ACE-FTS observations.
Both the integration or summation in the spatial and spectral domain could lead to a cancellation or compensation of model errors.
Furthermore, model spectra would be computed for all molecules known to be relevant for Earth's atmosphere, the impact of individual molecules is of little interest.
Finally, the state of the atmosphere should be known precisely; sophisticated closure experiments have been performed for this purpose \citep[e.g.][]{Turner04,Strow06,Masiello12,Oreopoulos12}.
Note that there appears to be no direct link between the ACE climatology \citep{Jones12,Koo17} and the ACE-FTS atlas \citep{Hughes14}.


\section{Summary and Outlook}
\label{sec:conclusions}

Effective height transit spectra of Earth have been generated by combining representative limb transmission spectra observed by the ACE-FTS solar occultation instrument.
These spectra have been degraded to moderate and low resolution and compared with spectra computed with an lbl radiative transfer code using HITRAN (or GEISA) spectroscopic data.
Inclusion of exclusion of molecules considered in the modeling allowed to study their impact on the transit spectra.
The main infrared absorbers water, carbon dioxide, ozone, nitrous oxide, and methane can be clearly identified in the effective height spectra.
Furthermore, nitric acid is very prominent around $900\cm$, and the main constituents of Earth's atmosphere, molecular oxygen and nitrogen, are also important for modeling the spectra. 
To further reduce the discrepancies, heavy molecules had to be considered, too.
In particular, the ``technosignatures'' CFC11 and CFC12 are visible in the moderate and low resolution spectra.

Transit observations of extrasolar Earth-like planets available for analysis in the near future will likely not encompass the large spectral range nor will they have the high resolution and low noise of the ACE-FTS spectra used in this study.
When ACE-FTS observations are used for a feasibility study of Earth-like exoplanet transit spectroscopy, the impact of spectral interval, resolution, and signal-noise ratio will be an important aspect.
Despite the clear visibility of a dozen or more molecules in the moderate and low resolution spectra, our SNR estimates indicate that only a few species may actually be detectable under very favorable conditions.

For the quantitative estimation of atmospheric state parameters from spectroscopic observations, inversion by numerical optimization techniques is well established for Earth and Solar System planets.
More recently, these techniques have also been applied successfully for remote sensing of exoplanets \citep[e.g.][]{Madhusudhan09,Lee12,Line12,Waldmann15t,Waldmann15e}.
So far these retrievals are confined mostly to large objects such as hot Jupiters, whereas the analysis of smaller objects such as super-Earths and Earth-like exoplanets is clearly more challenging \citep{Benneke12,Barstow13g}.
Nevertheless, we are adapting our GARLIC code (already used for analysis of microwave, far, thermal, and near IR Earth observation data \citep[e.g.][]{GimenoGarcia11,Xu16}) to exoplanet studies.

Verification and validation of exoplanet retrieval codes is an important aspect.
Whereas verification can be readily accomplished using synthetic measurements and code intercomparison (similar to \citep[e.g.][]{Clarmann03}), validation is challenging due to the lack of reference ``truth'' data (e.g.\ in situ measurements).
Thus, testing an exoplanet retrieval with Earth (or other Solar System planets such as Mars or Venus) is an attractive alternative.
The spectra provided in the ACE-FTS atlases provide an unique opportunity to generate a representative effective height spectrum (similar to the ATMOS spectra \citep{Abrams96a}), and our inverse problem solver currently being developed on the basis of the GARLIC forward model will be validated against this dataset.



\section*{Acknowledgments}
First we would like to thank Thomas Trautmann and Adrian Doicu (Ober\-pfaffenhofen) and Lee Grenfell and Heike Rauer (Berlin) for useful discussions and critical reading of the manuscript.
The atmospheric spectra provided by the MIPAS group in Oxford have been quite useful for this study, see \url{http://eodg.atm.ox.ac.uk/ATLAS/}.
Financial support by the Deutsche Forschungsgemeinschaft --- DFG (projects SCHR 1125/3-1 and GO 2610/1-1) is gratefully acknowledged.


\end{document}